\documentclass[11pt]{article}
\usepackage{amsfonts,amsmath,enumerate}

\usepackage{graphicx} 

\newtheorem{lemma}{Lemma}
\newtheorem{theorem}{Theorem}
\newtheorem{corollary}{Corollary}
\newtheorem{proposition}{Proposition}

\def\R{{\mathbb R}}
\def\C{{\mathbb C}}

\def\H{{\cal H}}

\def\beq{\begin{equation}}
\def\eeq{\end{equation}}

\begin{document}

\title{Two-dimensional rational solitons and their blow-up
via the Moutard transformation\thanks{This research was supported by
RFBR (grants 06-01-72551 (I.A.T.) and 06-01-00814 (S.P.T.)). The
first author (I.A.T.) was also supported by SB RAS (complex
integration project 2.15) and by Max Plank Instit\"ut f\"ur
Mathematik in Bonn.}}
\author{I.A. Taimanov
\thanks{Institute of Mathematics, 630090 Novosibirsk, Russia;
e-mail: \texttt{taimanov@math.nsc.ru}} \and S.P. Tsarev
\thanks{Krasnoyarsk State Pedagogical University,
ul. Levedevoi, 89, 660049
Krasnoyarsk, Russia; e-mail: \texttt{sptsarev@mail.ru}.}}

\date{}

\maketitle


\section{Introduction}

This article deals with applications of the Moutard transformation
\cite{Moutard} which is a two-dimensional version of the
well-known in solitonics Darboux transformation to some
problems of the spectral theory of two-dimensional operators and of
$(2+1)$-dimensional nonlinear evolution equations.

In particular, the main results consist in the explicit construction of

\begin{itemize}
\item
{\sl two-dimensional Schr\"odinger operators
$$
H = -\Delta + u= -(\partial_x^2 +\partial_y^2) + u(x,y)
$$
with fast decaying smooth rational potentials such that their $L_2$-ker\-nels
are nontrivial and, moreover, contain at least two-dimensional subspaces
spanned by rational eigenfunctions (see Theorems \ref{t-ord2} and
\ref{t-ord3});}

\item
{\sl blow-up  of solutions to the Novikov--Veselov (NV) equation,
which is a two-dimen\-sio\-nal generalization of the Korteweg--de
Vries (KdV) equation, with fast decaying rational Cauchy data (see
Theorem \ref{t-blowup}).}
\end{itemize}

The first construction  was already announced and briefly sketched in
\cite{TT}. For operators with such fast decaying potentials there exists
a nice spectral theory \cite{Faddeev,NK}.
We note that for one-dimensional Schr\"odinger
operators with fast decaying potentials the existence of
square-summable eigenfunctions at zero energy level is impossible
(see, for instance \cite{Faddeev}) and for higher-dimensional operators
(i.e., for dimensions greater or equal than $5$) one may easily construct such
examples for which the kernel contains a smoothing of the Green function
of $\Delta$. However for two-dimensional operators
these are the first known examples with nontrivial $L_2$-kernel.

The Novikov--Veselov equation has the form
\begin{equation}
\label{nv}
\begin{split}
U_t = \partial^3 U + \bar{\partial}^3 U + 3\partial(VU) +
3\bar{\partial} (\bar{V}U) =0,
\\
\bar{\partial}V = \partial U,
\end{split}
\end{equation}
and it is the first in the hierarchy of equations which have the form
$$
H_t = HA+BH,
$$
where $A$ and $B$ are differential operators. Here and everywhere
below we use the standard derivatives
$\partial=\partial_z=\frac{1}{2}(\partial_x- i \partial_y)$,
 $\bar\partial=\partial_{\bar z}=\frac{1}{2}(\partial_x + i \partial_y)$
w.r.t. $z=x+i\,y$. These equations were introduced in \cite{NV2} as
equations which preserve the zero-level spectrum of $H$. The
principal part of the $n$-th equation from the NV hierarchy takes
the form
$$
U_t = \partial^{2n+1}U  + \bar{\partial}^{2n+1} U + \dots
$$
where $\dots$ stays for terms of lower order.

It follows from the inverse scattering method that solutions to
the KdV equation with analytical fast decaying Cauchy data do not blow up
and the well-posedness of the Cauchy problem for this equation is established
in many functional spaces (see \cite{KPV} and references therein).
However as we show this is not true for its natural two-dimensional
generalization.

The examples of blow-up solutions were   obtained not by using
the inverse scattering method which is not well-developed in this situation.
Indeed, the inverse spectral problem for a two-dimensional Schr\"odinger
operator at a fixed energy level was first posed in \cite{DKN} and has been
studied for positive energy levels \cite{GM} or levels below the ground state
\cite{GN}. We think that the study of this problem on the zero energy level
will help to understand both phenomena which we discuss in this article.

We also have to remark that our potentials obtained by iterations of
the Moutard transformation may be considered as two-dimensional generalizations
of one-dimensional rational solitons obtained in the same way
using the Darboux transformation (see \S \ref{ss2-1}). However in the
one-dimensional case these potentials are always singular.

Finally we remark that our potentials are constructed by iterations
of the Moutard transformation from the constant potential and all
such potentials are integrable on the zero energy level in the sense
that all solutions to the equation $H\psi =0$ may be explicitly
constructed via quadratures from linear combinations of harmonic
functions. We give the details of this construction in
Section~\ref{sec-cube}.

The authors thank P.G. Grinevich and S.P. Novikov for useful discussions.

\section{Darboux and Moutard transformations}
\label{sec-dm}

\subsection{The Darboux transformation}
Let
$$
H = -\frac{d^2}{dx^2} + u(x)
$$
be a one-dimensional Schr\"odinger operator and let $\omega$
satisfy the equation
$$
H \omega = 0.
$$
The function $\omega$ determines a
factorization of $H$:
\beq
\label{onefactor}
H = A^\top A, \ \ \ A = -\frac{d}{dx} + v, \ \ \
A^\top = \frac{d}{dx}+v, \ \ \ v = \frac{\omega_x}{\omega}.
\eeq
Indeed we have
$$
A^\top A = \left(\frac{d}{dx} + v\right)\left(-\frac{d}{dx} +
v\right) = -\frac{d^2}{dx^2} + v^2 + v_x
$$
and the equation
$$
v_x + v^2 = u
$$
is equivalent to $H\omega = 0$. If
$v$ is real-valued we have $ A^\ast = A^\top$.

{\it The Darboux transformation} of $H$ \cite{Darboux}
is the swapping of $A^\top$
and $A$:
$$
H = A^\top A \to \widetilde{H} = AA^\top,
$$
or in terms of $u$:
$$
u = v^2 + v_x  \to \widetilde{u} = v^2 - v_x.
$$

It is easy to check the following:

\begin{proposition}
\label{prop1}
If $\varphi$ satisfies the equation
$H \varphi = E \varphi$ with $E = \mathrm{const}$
then $\widetilde{\varphi} = A \varphi$ satisfies the equation
$\widetilde{H} \widetilde{\varphi} = E \widetilde{\varphi}.$
\end{proposition}

{\sc Remark 1.}
In general the Darboux transformation is defined for
any solution to the equation $H \omega = c\omega$ with
$c=\mathrm{const}$.
In this case it reduces to the transformation
of $H^\prime = H-c$ for which $H^\prime\omega = 0$.

\subsection{One-dimensional solitons via the Darboux transformation}
\label{ss2-1}

Let $u =0$ and $\omega = \omega_1 = x = \tau_1$. Then
$$
v = \frac{1}{x},
\ \ \ v_x = - \frac{1}{x^2},
\ \ \ u_1 = \widetilde{u} = \frac{2}{x^2}.
$$
The function
$$
\psi(P,x) = \left(1 - \frac{1}{i\sqrt{E}x}\right)e^{i\sqrt{E}x}
$$
is meromorphic in $P=(E,\lambda)$
on the Riemann surface  $\Gamma = \{\lambda^2 = E\}$
and for every $E$ its branches give a basis of solutions to the equation
$$
H_1\psi = \left(-\frac{d^2}{dx^2} + u_1 \right)\psi = E \psi
$$
(we normalize $\psi$ by the condition $\psi \approx e^{i\sqrt{E}x}$ as
$E \to \infty$).
Now we may apply the Darboux transformation defined by
$\omega_2 = x^2 + \frac{\tau_2}{\omega_1}$ and obtain another potential.
In particular, for every $n$ the potential
$$
U_n = \frac{n(n+1)}{x^2},
$$
is obtained after $n$ iterations.
The orbit of $U_n$ under the Korteweg--de Vries hierarchy is the
$n$-dimensional family $M_n$ of potentials.

We have \cite{AM,AMM}:

\begin{enumerate}
\item
there is a general recursion procedure found by Adler and Moser
\cite{AM} for deriving the polynomials
$$
\theta_n(\tau_1,\dots,\tau_n)
$$
such that

\begin{enumerate}[(a)]
\item
$\theta_n$ is a polynomial of degree $\frac{n(n+1)}{2}$ in $x = \tau_1$.
In particular, we have
$$
\theta_1 = x = \tau_1,
$$
$$
\theta_2 = x^3 + \tau_2,
$$
$$
\theta_3 = x^6 + 5\tau_2 x^3 + \tau_3 x - 5 \tau_2^2;
$$

\item
the function
$$
u_n(x) = -2 \frac{d^2}{dx^2} \log \theta_n(\tau_1+x,\tau_2,\dots,\tau_n)
$$
is the potential of $H$ obtained after $n$ iterations of
the Darboux transformation (which starts at $u_0=0$).
Here $\tau_k, k =2,\dots$, are free scalar parameters (integration
constants) which appear one by one at every step of the iteration;

\item
for $\varphi_n = \frac{\theta_{n+1}}{\theta_n}$ we have
$$
\left(-\frac{d^2}{dx^2} + u_n\right)\varphi_n = 0;
$$
\end{enumerate}

\item
$M_n$ is an $n$-dimensional family parameterized by $\tau \in \C^n$
and consists of the potentials
$u_n(\tau_1+x,\tau_2,\dots,\tau_n$);

\item
there are birational transformations $\tau \to t$
of the form
$$
t_ k = a_k \tau_k + g_k(\tau_1,\dots,\tau_{k-1}), \ \ a_k \neq 0,
$$
such that
$$
\frac{\partial u_n}{\partial t_k} = X_k(u_n)
$$
is the $k$-th KdV flow;

\item
the zeroes $x_1,\dots,x_{n(n+1)/2}$ of the polynomials $\theta_n$
are evolved by the KdV-flows as some integrable Hamiltonian systems
and, for instance, for the original KdV equation
$$
u_t = 3uu_x - \frac{1}{2}u_{xxx} = X_2(u)
$$
their dynamics is described by the Calogero-Moser system.
\end{enumerate}

\subsection{The Moutard transformation}

Let $H$ be a two-dimensional potential Schr\"odinger operator
and let $\omega$ be a solution to the equation
$$
H \omega = (- \Delta + u )\omega = 0.
$$
Then {\it the (elliptic) Moutard transformation} of $H$ (cf.\ \cite{Moutard})
is defined as
$$
\widetilde{H} = -\Delta + u -2\Delta \log \omega =
-\Delta - u + 2\frac{\omega_x^2+\omega_y^2}{\omega^2}.
$$

\begin{proposition}
\label{prop3}
If $\varphi$ satisfies the equation
$H \varphi = 0$,
then the function $\theta$ defined from the system
\beq
\label{eigen-moutard}
(\omega \theta)_x = -\omega^2 \left(\frac{\varphi}{\omega}\right)_y, \ \ \
(\omega \theta)_y = \omega^2 \left(\frac{\varphi}{\omega}\right)_x
\eeq
satisfies
$\widetilde{H}\theta = 0$.
\end{proposition}

Obviously, if $\theta$ satisfies (\ref{eigen-moutard}) then
\beq
\label{theta-family}
\theta + \frac{C}{\omega}, \ \ \ C = \mathrm{const},
\eeq
satisfies (\ref{eigen-moutard}) for any constant  $C$.

We shall use the following notation for the Moutard transformation:
$$
M_\omega(u) = \widetilde{u} = u -2\Delta \log \omega, \ \ \
M_\omega(\varphi) = \{\theta+\frac{C}{\omega}, \ C \in \C\}.
$$

In the one-dimensional limit the Moutard transformation
reduces to the Darboux transformation. Indeed, let
$u=u(x)$ depend on $x$ only and
$\omega=f(x)e^{\sqrt{c}y}$.
Then $f$ satisfies the one-dimensional Schr\"odinger equation
$$
H_0 f = \left( - \frac{d^2}{dx^2} + u\right) f = cf
$$
and the Moutard transformation reduces to the Darboux transformation
of $H_0$ defined by $f$:
$$
H = H_0  - \frac{\partial^2}{\partial y^2} \ \ \longrightarrow \ \
\widetilde{H} =
\widetilde{H_0} -\frac{\partial^2}{\partial y^2}.
$$
If $g=g(x)$ satisfies $H_0 g = E g$, then
$H \varphi = 0$ with $\varphi = e^{\sqrt{E}y}g(x)$.
We derive from (\ref{eigen-moutard}) that
$\theta = e^{\sqrt{E}y}h(x)$ satisfies $\widetilde{H}\theta = 0$ if
$h = \frac{1}{\sqrt{c}+\sqrt{E}} \left(\frac{d}{dx} - \frac{f_x}{f}\right)g$,
i.e. $h$ is a multiple of the Darboux transform of $g$:
$h =  - \frac{1}{\sqrt{c}+\sqrt{E}}Ag$
where $H_0 - c = A^\top A$ is the factorization of $H_0-c$
defined by $f$.
The inverse Darboux transformation is given by
$g = \frac{1}{\sqrt{c}-\sqrt{E}}\left( \frac{d}{dx} + \frac{f_x}{f}\right)h$.

{\sc Remark 2.}
There is another two-dimensional generalization of the Darboux transformation
called the Laplace transformation. It is defined for a more general operator,
i.e. for the Schr\"odinger operator with an electromagnetic field:
$$
H = 4 (\bar{\partial} + \beta)(-\partial + \alpha) + u,
$$
and has the form
$$
H \to \widetilde{H} = 4u(-\partial +
\alpha)u^{-1}(\bar{\partial} + \beta) + u.
$$
So if $H\varphi=0$, then $\tilde{\varphi} =
(-\partial+\alpha)\varphi$ satisfies the equation
$\widetilde{H}\widetilde{\varphi}=0$. In the one-dimensional limit
it also reduces to the Darboux transformation.
Its relation to integrable systems was
recently studied in \cite{NV}.

\section{Soliton potentials via the Moutard transformation}

In \cite{TT} we gave simple examples of fast decaying
rational potentials of the two-dimensional Schr\"odinger operator
which have a degenerate $L_2$-kernel. These examples are
constructed  using the Moutard transformation as follows.

{\sc Main construction.}
Let
$$
H_0 = - \Delta = -\Delta + u_0
$$
be an operator with a potential $u_0 = 0$ and let $\omega_1$ and $\omega_2$
satisfy the equations
$$
H_0 \omega_1 = H_0 \omega_2 = 0.
$$
We take the Moutard transformations $M_{\omega_1}$ and $M_{\omega_2}$
defined by $\omega_1$ and $\omega_2$ and obtain the operators
$$
H_1 = -\Delta + u_1, \ \ \ H_2 = -\Delta + u_2
$$
where $u_1 = M_{\omega_1}(u_0), u_2 = M_{\omega_2}(u_0)$.
By the construction, we have
$$
H_1 M_{\omega_1}(\omega_2) = 0, \ \ \
H_2 M_{\omega_2}(\omega_1) = 0.
$$
Let us choose some function
$$
\theta_1 \in M_{\omega_1}(\omega_2)
$$
and put
$$
\theta_2  = -\frac{\omega_1}{\omega_2} \theta_1 \in M_{\omega_2}(\omega_1).
$$
These functions define the Moutard transformations of $H_1$ and $H_2$
and we obtain the operators $H_{12}$ and $H_{21}$ with the potentials
$$
u_{12} = M_{\theta_1}(u_1), \ \ \ \
u_{21} = M_{\theta_2}(u_2).
$$
The following classical lemma is checked by a
straightforward computation which
we omit.

\begin{lemma}
\begin{enumerate}
\item
$u_{12}=u_{21} = u$, i.e. the diagram
$$
\begin{array}{ccccc}
& u_0 & \stackrel{\omega_1}{\longrightarrow} & u_1 & \\
\omega_2 & \downarrow & & \downarrow & \theta_1 \\
& u_2 & \stackrel{\theta_2}
{\longrightarrow} & u_{12} = u_{21}&
\end{array},
$$
where $\theta_1 \in M_{\omega_1}(\omega_2), \ \theta_2 =
-\frac{\omega_1}{\omega_2}\theta_1 \in M_{\omega_2}(\omega_1)$,
is commutative;

\item
For $\psi_1 = \frac{1}{\theta_1}$ and $\psi_2 = \frac{1}{\theta_2}$ we have
$$
H\psi_1 = H\psi_2 = 0
$$
where $H = -\Delta+u$.
\end{enumerate}
\end{lemma}

We note that in this construction we have a free scalar parameter
$C$ (see (\ref{theta-family})) for the choice of $\theta_1 \in
M_{\omega_1}(\omega_2)$. This parameter can be used in some cases to
build a non-singular potential $u$ and functions $\psi_1$ and
$\psi_2$.

Now let us apply this construction to the situation when
$$
u_0 = 0.
$$
The desired formula for the potential $u$ is given by the following
theorem which is derived by simple and straightforward
computations which we omit.

\begin{theorem}
\label{p-iteration}
Let
$$
\omega_1 = p_1(z)+\overline{p_1(z)}, \ \ \
\omega_2 = p_2(z) + \overline{p_2(z)},
$$
where $p_1$ and $p_2$ are holomorphic functions of $z$. Let us
consider the Moutard transformation of the operator $H_0 = - \Delta$
defined by $\omega_1$. Then the corresponding transformation
(\ref{eigen-moutard}) of $\omega_2$
 has the form
$$
\theta_1 = \frac{i}{p_1+\bar{p}_1} \left(
(p_1\bar{p}_2-p_2\bar{p}_1) + \int\left((p^\prime_1 p_2 - p_1
p^\prime_2)dz + (\bar{p}_1\bar{p}^\prime_2 - \bar{p}^\prime_1
\bar{p}_2) d\bar{z}\right)\right).
$$
Therefore the second iteration defined by $\theta_1$ gives us the
operator $H = -\Delta  + u$ with
$$
u = -
 2 \Delta \log \omega_1 - 2\Delta \log\theta_1 =
 - 2 \Delta \log(\omega_1\theta_1) =
$$
\begin{equation}
\label{iteration}
- 2\Delta \log
i\left(
(p_1\bar{p}_2-p_2\bar{p}_1) +
\int\left((p^\prime_1 p_2 - p_1 p^\prime_2)dz + (\bar{p}_1\bar{p}^\prime_2 -
\bar{p}^\prime_1 \bar{p}_2) d\bar{z}
\right)\right).
\end{equation}
\end{theorem}

The free scalar parameter which we mentioned above appears as
the integration constant in (\ref{iteration}).

Let us apply this theorem to obtain some interesting examples of operators.
We consider hereafter only the cases when $\omega_1$ and $\omega_2$ are
real-valued harmonic polynomials.

{\sc Example 1}. Let
$$
\omega_1 =  x +2(x^2-y^2)+xy, \ \ \
\omega_2 =  x+y+\frac{3}{2}(x^2-y^2)+5xy,
$$
which in terms of complex polynomials is written as
\begin{equation}\label{p2ord}
p_1(z) = \left(1 - \frac{i}{4}\right) z^2 + \frac{z}{2}, \ \ \
p_2(z)  = \frac{1}{4}(3-5i)z^2 + \frac{1-i}{2}z.
\end{equation}
For some appropriate  constant  $C$ in $\theta_1$ we obtain
\begin{equation}
\label{ord2-1} u = -\frac{5120    (1 + 8    x + 2y + 17 x^2 + 17
y^2)}{(160 + 4    x^2 + 4y^2 + 16    x^3  + 4    x^2y + 16    x
y^2  + 4    y^3 + 17(x^2+y^2)^2)^2} =
\end{equation}
$$
-\frac{5120|1+(4-i)z|^2}{(160+|z|^2|2+(4-i)z|^2)^2}
$$
and
\begin{equation}
\label{ord2-2}
\begin{split}
\psi_1 = \frac{x + 2    x^2 + x    y - 2    y^2}{160 + 4    x^2 +
4y^2 +  16    x^3  + 4    x^2    y + 16    x    y^2 + 4
y^3 + 17 (x^2+y^2)^2}, \\
\psi_2 = \frac{2    x + 2y + 3    x^2 + 10    x    y - 3    y^2}{160
+ 4    x^2 + 4y^2 +  16    x^3  + 4    x^2    y + 16    x    y^2 + 4
y^3 + 17 (x^2+y^2)^2}
\end{split}
\end{equation}
(here we simplify the expressions for $\psi_1$ and $\psi_2$ by
multiplying them by some constant). On page~\pageref{page-pic} we
put the graphs of this potential and the solutions $\psi_1$,
$\psi_2$.

\begin{theorem}[\cite{TT}]
\label{t-ord2}
The potential $u$ given by (\ref{ord2-1}) is smooth, rational,  and
decays like $1/r^6$ for $r \to \infty$.

The functions $\psi_1$ and $\psi_2$ given by (\ref{ord2-2}) are smooth,
rational, decay like $1/r^2$ for $r \to \infty$ and span a
two-dimensional space in the kernel of the operator $L = -\Delta +u:
L_2(\R^2) \to L_2(\R^2)$.
\end{theorem}

Here and in the next example we put $r = \sqrt{x^2+y^2}$.

{\sc Example 2.} Let us take for $\omega_1$ and
$\omega_2$ two harmonic polynomials of the third order:
$$
\omega_1 = x + \frac{x^2-y^2-3xy}{5}+2(-x^3-3x^2y+3xy^2+y^3),
$$
$$
\omega_2 = x+y
+\frac{x^2-y^2}{2}-\frac{xy}{5}-4(3x^2y-y^3)
$$
which in terms of complex polynomials take the form
\begin{equation}\label{p3ord}
p_1(z)\! = \!(i\!-\! 1)z^3\! + \!\left(\frac{1}{10}\! +\!
\frac{3}{20}i\!\right)z^2 \!\!+\! \frac{z}{2}, \
 p_2(z)\! =\! 2iz^3\!\! +\!
\left(\frac{1}{4}\!+\!\frac{i}{20}\right)z^2\!\! +\!
\frac{1\!-\!i}{2}z.
\end{equation}
Then the potential $u$ and the functions $\psi_1$ and $\psi_2$ take
the form
\begin{equation}
\label{ord3}
u = \frac{F_0(x,y)}{G(x,y)^2}, \ \ \ \psi_1 = \frac{F_1(x,y)}{G(x,y)}, \ \ \
\psi_2 = \frac{F_2(x,y)}{G(x,y)}
\end{equation}
with
$$
F_0(x,y) = -1280000(25+20x-287x^2+60x^3+1800x^4 -30y -600xy -300x^2y +
$$
$$
313y^2 +60xy^2 + 3600x^2y^2 -300y^3 + 1800y^4),
$$
$$
G(x,y) = 40000 + 100x^2 +40x^3 -387x^4 + 40x^5 +800x^6 -60x^2 y -
$$
$$
800x^3y -200 x^4y + 100y^2 + 40xy^2 + 26x^2y^2 + 80 x^3y^2 + 2400x^4y^2
-60y^3 -
$$
$$
800xy^3 -
400x^2y^3+  413y^4 + 40 xy^4 + 2400x^2y^4 -200y^5 + 800y^6,
$$
$$
F_1(x,y) = -10x -2x^2 + 20 x^3 + 6xy + 60 x^2 y + 2y^2 - 60 x
y^2 - 20 y^3,
$$
$$
F_2(x,y) =
-10 x - 5 x^2 - 10 y + 2 x y + 120 x^2 y + 5 y^2 - 40
y^3
$$
(as in the previous example we simplify the expressions for $\psi_1$
and $\psi_2$  multiplying them by some constant). On
pages~\pageref{page-pic}--\pageref{page-pic2} we put the graphs of
this potential $u$ and the solutions $\psi_1$, $\psi_2$.

\begin{theorem}[\cite{TT}]
\label{t-ord3}
The potential $u$ given by (\ref{ord3}) is smooth, rational,  and
decays like $1/r^8$ for $r \to \infty$.

The functions $\psi_1$ and $\psi_2$ given by (\ref{ord3}) are smooth,
rational, decay like $1/r^3$ for $r \to \infty$ and span a
two-dimensional space in the kernel of the operator $L = -\Delta +u:
L_2(\R^2) \to L_2(\R^2)$.
\end{theorem}

{\sc Remark 3.} We guess that for every $N > 0$ by applying this construction
to other harmonic polynomials one can construct smooth rational
potentials $u$ and the eigenfunctions $\psi_1$ and $\psi_2$ decaying faster
than $\frac{1}{r^N}$.

\section{Soliton equations}

\subsection{Explicit
solutions to the Novikov--Veselov equation}

For simplicity, in this section
we renormalize the Schr\"odinger operator as follows
\begin{equation}
\label{renorm}
H = \partial\bar{\partial} + U = \frac{1}{4}\Delta - \frac{u}{4}
\end{equation}
with the standard $\partial=\partial_z=\frac{1}{2}(\partial_x- i \partial_y)$,
$\bar\partial=\partial_{\bar z}=\frac{1}{2}(\partial_x + i \partial_y)$.

The Moutard transformation defined by a function $\omega$
takes the form
$$
U \to U + 2 \partial\bar{\partial} \log \omega
$$
and the transformation of eigenfunctions is given by the same formulas
(\ref{eigen-moutard}) which are rewritten in terms of a complex coordinate
as
$$
\left(\bar{\partial} + \frac{\omega_{\bar{z}}}{\omega}\right) \theta =
i \left(\bar{\partial} - \frac{\omega_{\bar{z}}}{\omega}\right) \varphi,
\ \ \
\left(\partial + \frac{\omega_z}{\omega}\right) \theta =
-i \left(\partial - \frac{\omega_z}{\omega}\right) \varphi,
$$
or
$$
\varphi \to \theta = \frac{i}{\omega} \int (\varphi\partial \omega -
\omega \partial \varphi)dz - (\varphi \bar{\partial} \omega - \omega
\bar{\partial}\varphi)d\bar{z}.
$$

Let us assume that the function $\varphi$ depends also on the time $t$
and satisfies the equations
\begin{equation}
\label{eq}
\begin{split}
H \varphi = 0, \\
\partial_t \varphi = (\partial^3 + \bar{\partial}^3 + 3V\partial +
3V^\ast\partial)\varphi
\end{split}
\end{equation}
where
$$
\bar{\partial}V = \partial U, \ \ \ \partial V^\ast = \bar{\partial} U.
$$
Then by straightforward computations we derive

\begin{proposition}
[The extended Moutard transformation] \footnote{ This theorem was
first formulated in \cite{MS}. However in \cite{MS} the third line
in the formula for the Moutard transformation $\varphi \to \theta$
was omitted and here we make this correction. Due to this fact the
formula (6.2.5) in \cite{MS} does not give a solution of the
Novikov-Veselov equation. Recently we learned that such a correction
was already done in \cite{HLL}. See also \cite{AN} for another
approach to the extended Moutard transformations.}
\label{covariance} The system (\ref{eq}) is invariant under the
extended Moutard transformation
\begin{equation}
\label{ext}
\begin{split}
\varphi \to \theta = \frac{i}{\omega} \int (\varphi\partial \omega -
\omega \partial \varphi)dz - (\varphi \bar{\partial} \omega - \omega
\bar{\partial}\varphi)d\bar{z} +
\\
[\varphi\partial^3\omega - \omega\partial^3\varphi +
\omega\bar{\partial}^3\varphi - \varphi\bar{\partial}^3\omega
+2(\partial^2\varphi\partial\omega - \partial\varphi\partial^2\omega) -
\\
 2(\bar{\partial}^2\varphi\bar{\partial}\omega -
\bar{\partial}\varphi\bar{\partial}^2\omega)
+ 3V(\varphi\partial\omega  - \omega\partial\varphi)
+ 3V^*(\omega\bar{\partial}\varphi   - \varphi\bar{\partial}\omega)]dt,
\\
U \to U + 2\partial\bar{\partial}\log \omega, \ \
V \to V + 2 \partial^2 \log \omega, \ \
V^\ast \to V^\ast + 2 \bar{\partial}^2 \log \omega.
\end{split}
\end{equation}
If $\omega$ is a real-valued function, then the latter transformations
preserve the property $V^\ast = \overline{V}$.
\end{proposition}

The compatibility condition for the system (\ref{eq}) is
the system
$$
U_t = \partial^3 U + \bar{\partial}^3 U + 3\partial(VU) +
3\bar{\partial} (V^\ast U) =0,
$$
$$
\bar{\partial}V = \partial U, \ \ \ \partial V^\ast = \bar{\partial} U
$$
which for
$$
V^\ast = \overline{V}
$$
reduces to the Novikov--Veselov equation. In the sequel, we consider only the
cases when the above equality holds.

Let us consider the flow on the space of functions $p(z,t)$ holomorphic in $z$:
\begin{equation}
\label{flow}
\frac{\partial p}{\partial t} = \frac{\partial^3 p}{\partial z}.
\end{equation}

If $p_1(z,t)$ and $p_2(z,t)$ satisfy (\ref{flow}) then the functions
$\omega = p_1 + \bar{p}_1$ and $\varphi = p_2 + \bar{p}_2$ satisfy
the system (\ref{eq}) with $U = V = 0$. The extended Moutard
transformation differs from the original one by the
$dt$-term:
$$
\theta = \frac{i}{\omega} \int\Big(\Psi_1(\omega,\varphi)dz +
\Psi_2(\omega,\varphi)d\bar z + \Theta(\omega,\varphi)dt\Big).
$$
Let us substitute $\omega = p_1+\bar{p}_1$, $\varphi = p_2 +
\bar{p}_2$ into $\Psi_s$. By (\ref{flow}) and $V=V^\ast=0$ the
integrand is a closed form which implies
$$
\frac{\partial \Psi_1(p_1+\bar{p}_1,p_2+\bar{p}_2)}{\partial t} =
\frac{\partial \Theta(p_1+\bar{p}_1,p_2+\bar{p}_2)}{\partial z},
$$
$$
\frac{\partial \Psi_2(p_1+\bar{p}_1,p_2+\bar{p}_2)}{\partial t} =
\frac{\partial \Theta(p_1+\bar{p}_1,p_2+\bar{p}_2)}{\partial \bar
z}.
$$
Hence $\Theta_z$ and $\Theta_{\bar{z}}$ are defined by $\Psi_s$ and
we have the following analog of Theorem \ref{p-iteration}:

\begin{proposition}
\label{ext-iteration}
Let
$$
\omega_1 = p_1(z,t)+\overline{p_1(z,t)}, \ \ \ \omega_2 = p_2(z,t) +
\overline{p_2(z,t)},
$$
where $p_1$ and $p_2$ are holomorphic functions of $z$ which satisfy
(\ref{flow}). Let us consider the extended Moutard transformation of
the operator $H_0 = \partial\bar{\partial}$ defined by $\omega_1$,
and let $\theta_1$ be the image of $\omega_2$ under this
transformation. Let $H = \partial\bar{\partial}  + U$ is obtained by
the iteration of the Moutard transformation defined by $\theta_1$.
Then
\begin{equation}
\label{iteration2}
\begin{split}
U(z,\bar{z},t) = 2\partial\bar{\partial} \log i \Big(
(p_1\bar{p_2}-p_2\bar{p_1}) + \int\left((p_1^\prime p_2 - p_1
p_2^\prime)dz + (\bar{p_1}\bar{p_2}^\prime - \bar{p_1}^\prime
\bar{p_2}) d\bar{z}\right) +
\\
+ \int (p_1^{\prime\prime\prime}p_2 - p_1p_2^{\prime\prime\prime} +
2(p_1^\prime p_2^{\prime\prime} - p_1^{\prime\prime}p_2^\prime) +
\bar{p_1}\bar{p_2}^{\prime\prime\prime} -
\bar{p_1}^{\prime\prime\prime}\bar{p_2} +
2(\bar{p_1}^{\prime\prime}\bar{p_2}^\prime -
\bar{p_1}^\prime\bar{p_2}^{\prime\prime}))dt \Big).
\end{split}
\end{equation}
\end{proposition}

\begin{corollary}
\label{cor1}
Let $p_1(z,t)$ and $p_2(z,t)$ be holomorphic functions
in $z$ which satisfy the equation (\ref{flow}).

Then  the substitution of  them into (\ref{iteration2}) gives a
solution to the Novikov--Veselov equation, which is rational in
$z,\bar{z}$, and $t$.
\end{corollary}

{\sc Remark 4.} It is clear that the analogs of (\ref{ext}) may be
derived for all equations from the Novikov--Veselov hierarchy and
hence the explicit solutions to them, in particular, of the form
(\ref{ext}) may be constructed.

\subsection{$\sigma$-flows}

Let us denote by $\H_N$ the space of all harmonic polynomials
$p(z)+\overline{p(z)}$ of the form
$$
p(z) = \sigma_0 z^N + \sigma_1 z^{N-1} + \dots + \sigma_{N-1} z + \sigma_N
$$
and denote by $\H^0_N$ the subspace $\sigma_0=1$.

The  flow (\ref{flow})
generates on $\H_N$ a linear flow
\begin{equation}
\label{sigma-0}
\dot{\sigma}_k = (N-k+3)(N-k+2)(N-k+1) \sigma_{k-3}, \ \ \ k=0,\dots,N.
\end{equation}
Since $\sigma_0$ is constant along the flow we restrict these equations onto
$\H^0_N$ and obtain a particular example of the linear system on
$\sigma_1,\dots,\sigma_N$:
\begin{equation}
\label{sigma-1}
\dot{\sigma} = A + B \sigma
\end{equation}
which generates a dynamical system on the $n$-th
symmetric product $S^n\C$ of $\C$. Indeed, $\sigma_1,\dots,
\sigma_n$
are the elementary symmetric polynomials in the roots
$z_1,\dots,z_n$ of $p(z)$:
$$
\sigma_1(z_1,\dots,z_n) = -(z_1+\dots+z_n), \dots,
\sigma_n(z_1,\dots,z_n) = (-1)^{n}z_1\dots z_n
$$
and the integrable (even linear) evolution of $\sigma$ induces a dynamical
system on $S^n\C$. We call such a dynamical system on $S^n\C$ a
$\sigma$-system.

In contrast to the Calogero--Moser systems,
$\sigma$-systems do not describe the evolution of singularities of a
solution to the NV equation (as the Calogero--Moser flow does for
the KdV equation), i.e. the evolution of particle-type solutions \cite{DN,AMM}
(see also \S \ref{ss2-1}). In fact,

\begin{itemize}
\item
{\sl the Calogero--Moser systems and $\sigma$-systems are
different in many respects.}
\end{itemize}

For example, let us consider the simplest flow induced by
(\ref{flow}) on $\H_3$:
$$
\dot{\sigma}_1=0, \ \ \dot{\sigma}_2 = 0, \ \ \dot{\sigma}_3 = 6.
$$
The general solution is
$$
\sigma_1 = a_1, \ \ \sigma_2 = a_2, \ \ \sigma_3 = a_3 +6t
$$
However for a solution $(x_1(t),\dots,x_n(t))$ to the Calogero--Moser system
the $k$-th elementary symmetric polynomial $\sigma_k(x_1,\dots,x_n)$ is
a polynomial in $t$ of degree $k$ and especially any function $x_k(t)$ is
algebraic (see Corollary 3 on p. 118 in \cite{AMM}).

So we may present examples when the solutions to the Calogero--Moser
system and to a $\sigma$-system are related by a reparameterization
of the time variable. For instance, such is the equilateral
triangular solution  to the Calogero--Moser system which corresponds
to the solution
\\
$-2\frac{d^2}{dx^2}\left[(x-x_1(t))(x-x_2(t)) (x-x_3(t))\right]$ of
the KdV equation:
$$
x_k(t) = \varepsilon^{k} \sqrt[3]{\varepsilon}t, \ \ \varepsilon =
e^{\frac{2\pi i}{3}}, \ \ k=1,2,3;
$$
and a solution to the $\sigma$-system corresponding to $p(z,t) =
z^3 + 6t$:
$$
z_k(t) = \varepsilon^k \sqrt[3]{6t}, \ \ \varepsilon =
e^{\frac{2\pi i}{3}}, \ \ k=1,2,3.
$$

By Corollary \ref{cor1}, we have

\begin{itemize}
\item
Given two solutions $p_1 \in \H^0_M$ and $p_2 \in \H^0_N$ of
(\ref{sigma-0}), by a substitution of $e^{i\lambda_1}p_1$ and
$e^{i\lambda_2}p_2$, where $\lambda_1$ and $\lambda_2$ are
real-valued constants, into (\ref{iteration2}) we obtain solutions
of the Novikov--Veselov equation.\footnote{If we multiply, for
instance, $p_1$ by a constant $\mu \in \R$ then $\omega = p_1 +
\bar{p_1}$ is multiplied by $\mu$ and, since the Moutard
transformation depends on $\omega$ taken up to a multiple, this does
not change the transformation.} Therefore to each pair of solutions
to (\ref{sigma-0}) there corresponds an $(S^1 \times S^1)$-family of
solutions to the Novikov--Veselov equation.
\end{itemize}

We think that due to their natural appearance and not only because of their
relation to the Novikov--Veselov equation which we did establish,
$\sigma$-systems deserve a special investigation.

\subsection{On the blow-up of solutions}\label{sec-blow}

For simplicity let us write the formula (\ref{iteration2}) as
$$
U = 2\partial \bar{\partial} \log \Phi(p_1,p_2).
$$
(see the renomalization formula (\ref{renorm})). Due to the
integration in (\ref{iteration}) the function $\Phi(p_1,p_2)$ is
defined up to a constant which we take real-valued to obtain a
real-valued potential $U(z,\bar{z})$. By Proposition
\ref{covariance}, the potential $V = \bar{V}^\ast$ equals
$$
V = 2 \partial^2 \log \Phi(p_1,p_2).
$$

It is easy to check the following:
\begin{corollary}
The potential (\ref{ord2-1}) (see Theorem \ref{t-ord2}) is a
stationary solution of the Novikov--Veselov equation.
\end{corollary}

On the other hand one can obtain  solutions of the Novikov-Veselov
equations which blow up in finite time:

\begin{theorem}
\label{t-blowup} The solution $U(z,\bar{z},t)$ of the
Novikov--Veselov equation obtained form the polynomials $p_1=i\,
z^{2}$, $p_2= z^{2}+(1+i)z$ using Proposition~\ref{ext-iteration}
and the integration constant $C=-20$ in (\ref{ext}) has the form
$$U= \frac{H_1}{H_2},
$$
with $
  H_1=-12\Big(24 t x^2 + 12 t x + 24 t y^2 + 12 t y + x^5 - 3 x^4 y + 2 x^4
- 2 x^3 y^2 - 4 x^3 y - 2 x^2 y^3 - 60 x^2 - 3 x y^4 - 4 x  y^3 - 30
x + y^5 + 2 y^4 - 60 y^2 - 30 y\Big),
$
$$
H_2=(3 x^4 + 4 x^3 + 6 x^2 y^2 + 3 y^4 + 4 y^3 + 30 - 12 t)^2.
$$
This solution decays at infinity as $r^{-3}$ and obviously blows up
for some $t>0$.
\end{theorem}

\section{The cubic superposition formula}\label{sec-cube}

In this Section we extend the algebraic superposition formula of
L.~Bian\-chi~\cite{bianchi} for three initial solutions $\omega_1$,
$\omega_2$, $\omega_3$ of the Moutard equation to the case of the
Novikov-Veselov equation. In fact, already the extended Moutard
transformation (\ref{ext}) should be considered as an ``extended
superposition formula'' which produces a new solution $\theta$
starting with two solutions $\omega$, $\varphi$ of the linear
problem (\ref{eq}) with some initial potentials $U$, $V$, $V^*$. In
contrast to the well-known algebraic superposition formula for two
solutions of the Sine-Gordon equation, (\ref{ext}) is not algebraic
and requires a quadrature. This is typical for $(2+1)$-dimensional
B\"acklund transformations. In this case an algebraic superposition
formula exists for three initial solutions (cf.\ \cite{GTs96} for a
detailed discussion of this phenomenon).

\begin{figure}[htbp]
\begin{center}
 \unitlength 0.70mm \linethickness{0.4pt}
\begin{picture}(159.00,77.33)(0.0,70.0)
\multiput(102.00,147.33)(0.79,-0.09){2}{\line(1,0){0.79}}
\multiput(103.59,147.15)(0.30,-0.11){5}{\line(1,0){0.30}}
\multiput(105.09,146.61)(0.17,-0.11){8}{\line(1,0){0.17}}
\multiput(106.43,145.75)(0.11,-0.11){10}{\line(0,-1){0.11}}
\multiput(107.54,144.60)(0.12,-0.20){7}{\line(0,-1){0.20}}
\multiput(108.37,143.24)(0.10,-0.30){5}{\line(0,-1){0.30}}
\multiput(108.86,141.72)(0.07,-0.80){2}{\line(0,-1){0.80}}
\multiput(109.00,140.13)(-0.11,-0.79){2}{\line(0,-1){0.79}}
\multiput(108.77,138.55)(-0.12,-0.30){5}{\line(0,-1){0.30}}
\multiput(108.19,137.06)(-0.11,-0.16){8}{\line(0,-1){0.16}}
\multiput(107.29,135.75)(-0.13,-0.12){9}{\line(-1,0){0.13}}
\multiput(106.11,134.67)(-0.20,-0.11){7}{\line(-1,0){0.20}}
\multiput(104.72,133.88)(-0.38,-0.11){4}{\line(-1,0){0.38}}
\put(103.19,133.43){\line(-1,0){1.59}}
\multiput(101.60,133.34)(-0.52,0.09){3}{\line(-1,0){0.52}}
\multiput(100.03,133.61)(-0.25,0.10){6}{\line(-1,0){0.25}}
\multiput(98.56,134.23)(-0.16,0.12){8}{\line(-1,0){0.16}}
\multiput(97.27,135.17)(-0.12,0.13){9}{\line(0,1){0.13}}
\multiput(96.22,136.38)(-0.11,0.20){7}{\line(0,1){0.20}}
\multiput(95.48,137.79)(-0.10,0.39){4}{\line(0,1){0.39}}
\put(95.07,139.33){\line(0,1){1.60}}
\multiput(95.03,140.93)(0.11,0.52){3}{\line(0,1){0.52}}
\multiput(95.34,142.49)(0.11,0.24){6}{\line(0,1){0.24}}
\multiput(96.01,143.94)(0.11,0.14){9}{\line(0,1){0.14}}
\multiput(96.98,145.21)(0.14,0.11){9}{\line(1,0){0.14}}
\multiput(98.22,146.22)(0.24,0.12){6}{\line(1,0){0.24}}
\multiput(99.65,146.92)(0.59,0.10){4}{\line(1,0){0.59}}
\multiput(152.00,147.33)(0.79,-0.09){2}{\line(1,0){0.79}}
\multiput(153.59,147.15)(0.30,-0.11){5}{\line(1,0){0.30}}
\multiput(155.09,146.61)(0.17,-0.11){8}{\line(1,0){0.17}}
\multiput(156.43,145.75)(0.11,-0.11){10}{\line(0,-1){0.11}}
\multiput(157.54,144.60)(0.12,-0.20){7}{\line(0,-1){0.20}}
\multiput(158.37,143.24)(0.10,-0.30){5}{\line(0,-1){0.30}}
\multiput(158.86,141.72)(0.07,-0.80){2}{\line(0,-1){0.80}}
\multiput(159.00,140.13)(-0.11,-0.79){2}{\line(0,-1){0.79}}
\multiput(158.77,138.55)(-0.12,-0.30){5}{\line(0,-1){0.30}}
\multiput(158.19,137.06)(-0.11,-0.16){8}{\line(0,-1){0.16}}
\multiput(157.29,135.75)(-0.13,-0.12){9}{\line(-1,0){0.13}}
\multiput(156.11,134.67)(-0.20,-0.11){7}{\line(-1,0){0.20}}
\multiput(154.72,133.88)(-0.38,-0.11){4}{\line(-1,0){0.38}}
\put(153.19,133.43){\line(-1,0){1.59}}
\multiput(151.60,133.34)(-0.52,0.09){3}{\line(-1,0){0.52}}
\multiput(150.03,133.61)(-0.25,0.10){6}{\line(-1,0){0.25}}
\multiput(148.56,134.23)(-0.16,0.12){8}{\line(-1,0){0.16}}
\multiput(147.27,135.17)(-0.12,0.13){9}{\line(0,1){0.13}}
\multiput(146.22,136.38)(-0.11,0.20){7}{\line(0,1){0.20}}
\multiput(145.48,137.79)(-0.10,0.39){4}{\line(0,1){0.39}}
\put(145.07,139.33){\line(0,1){1.60}}
\multiput(145.03,140.93)(0.11,0.52){3}{\line(0,1){0.52}}
\multiput(145.34,142.49)(0.11,0.24){6}{\line(0,1){0.24}}
\multiput(146.01,143.94)(0.11,0.14){9}{\line(0,1){0.14}}
\multiput(146.98,145.21)(0.14,0.11){9}{\line(1,0){0.14}}
\multiput(148.22,146.22)(0.24,0.12){6}{\line(1,0){0.24}}
\multiput(149.65,146.92)(0.59,0.10){4}{\line(1,0){0.59}}
\multiput(132.00,127.33)(0.79,-0.09){2}{\line(1,0){0.79}}
\multiput(133.59,127.15)(0.30,-0.11){5}{\line(1,0){0.30}}
\multiput(135.09,126.61)(0.17,-0.11){8}{\line(1,0){0.17}}
\multiput(136.43,125.75)(0.11,-0.11){10}{\line(0,-1){0.11}}
\multiput(137.54,124.60)(0.12,-0.20){7}{\line(0,-1){0.20}}
\multiput(138.37,123.24)(0.10,-0.30){5}{\line(0,-1){0.30}}
\multiput(138.86,121.72)(0.07,-0.80){2}{\line(0,-1){0.80}}
\multiput(139.00,120.13)(-0.11,-0.79){2}{\line(0,-1){0.79}}
\multiput(138.77,118.55)(-0.12,-0.30){5}{\line(0,-1){0.30}}
\multiput(138.19,117.06)(-0.11,-0.16){8}{\line(0,-1){0.16}}
\multiput(137.29,115.75)(-0.13,-0.12){9}{\line(-1,0){0.13}}
\multiput(136.11,114.67)(-0.20,-0.11){7}{\line(-1,0){0.20}}
\multiput(134.72,113.88)(-0.38,-0.11){4}{\line(-1,0){0.38}}
\put(133.19,113.43){\line(-1,0){1.59}}
\multiput(131.60,113.34)(-0.52,0.09){3}{\line(-1,0){0.52}}
\multiput(130.03,113.61)(-0.25,0.10){6}{\line(-1,0){0.25}}
\multiput(128.56,114.23)(-0.16,0.12){8}{\line(-1,0){0.16}}
\multiput(127.27,115.17)(-0.12,0.13){9}{\line(0,1){0.13}}
\multiput(126.22,116.38)(-0.11,0.20){7}{\line(0,1){0.20}}
\multiput(125.48,117.79)(-0.10,0.39){4}{\line(0,1){0.39}}
\put(125.07,119.33){\line(0,1){1.60}}
\multiput(125.03,120.93)(0.11,0.52){3}{\line(0,1){0.52}}
\multiput(125.34,122.49)(0.11,0.24){6}{\line(0,1){0.24}}
\multiput(126.01,123.94)(0.11,0.14){9}{\line(0,1){0.14}}
\multiput(126.98,125.21)(0.14,0.11){9}{\line(1,0){0.14}}
\multiput(128.22,126.22)(0.24,0.12){6}{\line(1,0){0.24}}
\multiput(129.65,126.92)(0.59,0.10){4}{\line(1,0){0.59}}
\multiput(82.00,127.33)(0.79,-0.09){2}{\line(1,0){0.79}}
\multiput(83.59,127.15)(0.30,-0.11){5}{\line(1,0){0.30}}
\multiput(85.09,126.61)(0.17,-0.11){8}{\line(1,0){0.17}}
\multiput(86.43,125.75)(0.11,-0.11){10}{\line(0,-1){0.11}}
\multiput(87.54,124.60)(0.12,-0.20){7}{\line(0,-1){0.20}}
\multiput(88.37,123.24)(0.10,-0.30){5}{\line(0,-1){0.30}}
\multiput(88.86,121.72)(0.07,-0.80){2}{\line(0,-1){0.80}}
\multiput(89.00,120.13)(-0.11,-0.79){2}{\line(0,-1){0.79}}
\multiput(88.77,118.55)(-0.12,-0.30){5}{\line(0,-1){0.30}}
\multiput(88.19,117.06)(-0.11,-0.16){8}{\line(0,-1){0.16}}
\multiput(87.29,115.75)(-0.13,-0.12){9}{\line(-1,0){0.13}}
\multiput(86.11,114.67)(-0.20,-0.11){7}{\line(-1,0){0.20}}
\multiput(84.72,113.88)(-0.38,-0.11){4}{\line(-1,0){0.38}}
\put(83.19,113.43){\line(-1,0){1.59}}
\multiput(81.60,113.34)(-0.52,0.09){3}{\line(-1,0){0.52}}
\multiput(80.03,113.61)(-0.25,0.10){6}{\line(-1,0){0.25}}
\multiput(78.56,114.23)(-0.16,0.12){8}{\line(-1,0){0.16}}
\multiput(77.27,115.17)(-0.12,0.13){9}{\line(0,1){0.13}}
\multiput(76.22,116.38)(-0.11,0.20){7}{\line(0,1){0.20}}
\multiput(75.48,117.79)(-0.10,0.39){4}{\line(0,1){0.39}}
\put(75.07,119.33){\line(0,1){1.60}}
\multiput(75.03,120.93)(0.11,0.52){3}{\line(0,1){0.52}}
\multiput(75.34,122.49)(0.11,0.24){6}{\line(0,1){0.24}}
\multiput(76.01,123.94)(0.11,0.14){9}{\line(0,1){0.14}}
\multiput(76.98,125.21)(0.14,0.11){9}{\line(1,0){0.14}}
\multiput(78.22,126.22)(0.24,0.12){6}{\line(1,0){0.24}}
\multiput(79.65,126.92)(0.59,0.10){4}{\line(1,0){0.59}}
\multiput(82.00,77.33)(0.79,-0.09){2}{\line(1,0){0.79}}
\multiput(83.59,77.15)(0.30,-0.11){5}{\line(1,0){0.30}}
\multiput(85.09,76.61)(0.17,-0.11){8}{\line(1,0){0.17}}
\multiput(86.43,75.75)(0.11,-0.11){10}{\line(0,-1){0.11}}
\multiput(87.54,74.60)(0.12,-0.20){7}{\line(0,-1){0.20}}
\multiput(88.37,73.24)(0.10,-0.30){5}{\line(0,-1){0.30}}
\multiput(88.86,71.72)(0.07,-0.80){2}{\line(0,-1){0.80}}
\multiput(89.00,70.13)(-0.11,-0.79){2}{\line(0,-1){0.79}}
\multiput(88.77,68.55)(-0.12,-0.30){5}{\line(0,-1){0.30}}
\multiput(88.19,67.06)(-0.11,-0.16){8}{\line(0,-1){0.16}}
\multiput(87.29,65.75)(-0.13,-0.12){9}{\line(-1,0){0.13}}
\multiput(86.11,64.67)(-0.20,-0.11){7}{\line(-1,0){0.20}}
\multiput(84.72,63.88)(-0.38,-0.11){4}{\line(-1,0){0.38}}
\put(83.19,63.43){\line(-1,0){1.59}}
\multiput(81.60,63.34)(-0.52,0.09){3}{\line(-1,0){0.52}}
\multiput(80.03,63.61)(-0.25,0.10){6}{\line(-1,0){0.25}}
\multiput(78.56,64.23)(-0.16,0.12){8}{\line(-1,0){0.16}}
\multiput(77.27,65.17)(-0.12,0.13){9}{\line(0,1){0.13}}
\multiput(76.22,66.38)(-0.11,0.20){7}{\line(0,1){0.20}}
\multiput(75.48,67.79)(-0.10,0.39){4}{\line(0,1){0.39}}
\put(75.07,69.33){\line(0,1){1.60}}
\multiput(75.03,70.93)(0.11,0.52){3}{\line(0,1){0.52}}
\multiput(75.34,72.49)(0.11,0.24){6}{\line(0,1){0.24}}
\multiput(76.01,73.94)(0.11,0.14){9}{\line(0,1){0.14}}
\multiput(76.98,75.21)(0.14,0.11){9}{\line(1,0){0.14}}
\multiput(78.22,76.22)(0.24,0.12){6}{\line(1,0){0.24}}
\multiput(79.65,76.92)(0.59,0.10){4}{\line(1,0){0.59}}
\multiput(132.00,77.33)(0.79,-0.09){2}{\line(1,0){0.79}}
\multiput(133.59,77.15)(0.30,-0.11){5}{\line(1,0){0.30}}
\multiput(135.09,76.61)(0.17,-0.11){8}{\line(1,0){0.17}}
\multiput(136.43,75.75)(0.11,-0.11){10}{\line(0,-1){0.11}}
\multiput(137.54,74.60)(0.12,-0.20){7}{\line(0,-1){0.20}}
\multiput(138.37,73.24)(0.10,-0.30){5}{\line(0,-1){0.30}}
\multiput(138.86,71.72)(0.07,-0.80){2}{\line(0,-1){0.80}}
\multiput(139.00,70.13)(-0.11,-0.79){2}{\line(0,-1){0.79}}
\multiput(138.77,68.55)(-0.12,-0.30){5}{\line(0,-1){0.30}}
\multiput(138.19,67.06)(-0.11,-0.16){8}{\line(0,-1){0.16}}
\multiput(137.29,65.75)(-0.13,-0.12){9}{\line(-1,0){0.13}}
\multiput(136.11,64.67)(-0.20,-0.11){7}{\line(-1,0){0.20}}
\multiput(134.72,63.88)(-0.38,-0.11){4}{\line(-1,0){0.38}}
\put(133.19,63.43){\line(-1,0){1.59}}
\multiput(131.60,63.34)(-0.52,0.09){3}{\line(-1,0){0.52}}
\multiput(130.03,63.61)(-0.25,0.10){6}{\line(-1,0){0.25}}
\multiput(128.56,64.23)(-0.16,0.12){8}{\line(-1,0){0.16}}
\multiput(127.27,65.17)(-0.12,0.13){9}{\line(0,1){0.13}}
\multiput(126.22,66.38)(-0.11,0.20){7}{\line(0,1){0.20}}
\multiput(125.48,67.79)(-0.10,0.39){4}{\line(0,1){0.39}}
\put(125.07,69.33){\line(0,1){1.60}}
\multiput(125.03,70.93)(0.11,0.52){3}{\line(0,1){0.52}}
\multiput(125.34,72.49)(0.11,0.24){6}{\line(0,1){0.24}}
\multiput(126.01,73.94)(0.11,0.14){9}{\line(0,1){0.14}}
\multiput(126.98,75.21)(0.14,0.11){9}{\line(1,0){0.14}}
\multiput(128.22,76.22)(0.24,0.12){6}{\line(1,0){0.24}}
\multiput(129.65,76.92)(0.59,0.10){4}{\line(1,0){0.59}}
\multiput(102.00,97.33)(0.79,-0.09){2}{\line(1,0){0.79}}
\multiput(103.59,97.15)(0.30,-0.11){5}{\line(1,0){0.30}}
\multiput(105.09,96.61)(0.17,-0.11){8}{\line(1,0){0.17}}
\multiput(106.43,95.75)(0.11,-0.11){10}{\line(0,-1){0.11}}
\multiput(107.54,94.60)(0.12,-0.20){7}{\line(0,-1){0.20}}
\multiput(108.37,93.24)(0.10,-0.30){5}{\line(0,-1){0.30}}
\multiput(108.86,91.72)(0.07,-0.80){2}{\line(0,-1){0.80}}
\multiput(109.00,90.13)(-0.11,-0.79){2}{\line(0,-1){0.79}}
\multiput(108.77,88.55)(-0.12,-0.30){5}{\line(0,-1){0.30}}
\multiput(108.19,87.06)(-0.11,-0.16){8}{\line(0,-1){0.16}}
\multiput(107.29,85.75)(-0.13,-0.12){9}{\line(-1,0){0.13}}
\multiput(106.11,84.67)(-0.20,-0.11){7}{\line(-1,0){0.20}}
\multiput(104.72,83.88)(-0.38,-0.11){4}{\line(-1,0){0.38}}
\put(103.19,83.43){\line(-1,0){1.59}}
\multiput(101.60,83.34)(-0.52,0.09){3}{\line(-1,0){0.52}}
\multiput(100.03,83.61)(-0.25,0.10){6}{\line(-1,0){0.25}}
\multiput(98.56,84.23)(-0.16,0.12){8}{\line(-1,0){0.16}}
\multiput(97.27,85.17)(-0.12,0.13){9}{\line(0,1){0.13}}
\multiput(96.22,86.38)(-0.11,0.20){7}{\line(0,1){0.20}}
\multiput(95.48,87.79)(-0.10,0.39){4}{\line(0,1){0.39}}
\put(95.07,89.33){\line(0,1){1.60}}
\multiput(95.03,90.93)(0.11,0.52){3}{\line(0,1){0.52}}
\multiput(95.34,92.49)(0.11,0.24){6}{\line(0,1){0.24}}
\multiput(96.01,93.94)(0.11,0.14){9}{\line(0,1){0.14}}
\multiput(96.98,95.21)(0.14,0.11){9}{\line(1,0){0.14}}
\multiput(98.22,96.22)(0.24,0.12){6}{\line(1,0){0.24}}
\multiput(99.65,96.92)(0.59,0.10){4}{\line(1,0){0.59}}
\multiput(152.00,97.33)(0.79,-0.09){2}{\line(1,0){0.79}}
\multiput(153.59,97.15)(0.30,-0.11){5}{\line(1,0){0.30}}
\multiput(155.09,96.61)(0.17,-0.11){8}{\line(1,0){0.17}}
\multiput(156.43,95.75)(0.11,-0.11){10}{\line(0,-1){0.11}}
\multiput(157.54,94.60)(0.12,-0.20){7}{\line(0,-1){0.20}}
\multiput(158.37,93.24)(0.10,-0.30){5}{\line(0,-1){0.30}}
\multiput(158.86,91.72)(0.07,-0.80){2}{\line(0,-1){0.80}}
\multiput(159.00,90.13)(-0.11,-0.79){2}{\line(0,-1){0.79}}
\multiput(158.77,88.55)(-0.12,-0.30){5}{\line(0,-1){0.30}}
\multiput(158.19,87.06)(-0.11,-0.16){8}{\line(0,-1){0.16}}
\multiput(157.29,85.75)(-0.13,-0.12){9}{\line(-1,0){0.13}}
\multiput(156.11,84.67)(-0.20,-0.11){7}{\line(-1,0){0.20}}
\multiput(154.72,83.88)(-0.38,-0.11){4}{\line(-1,0){0.38}}
\put(153.19,83.43){\line(-1,0){1.59}}
\multiput(151.60,83.34)(-0.52,0.09){3}{\line(-1,0){0.52}}
\multiput(150.03,83.61)(-0.25,0.10){6}{\line(-1,0){0.25}}
\multiput(148.56,84.23)(-0.16,0.12){8}{\line(-1,0){0.16}}
\multiput(147.27,85.17)(-0.12,0.13){9}{\line(0,1){0.13}}
\multiput(146.22,86.38)(-0.11,0.20){7}{\line(0,1){0.20}}
\multiput(145.48,87.79)(-0.10,0.39){4}{\line(0,1){0.39}}
\put(145.07,89.33){\line(0,1){1.60}}
\multiput(145.03,90.93)(0.11,0.52){3}{\line(0,1){0.52}}
\multiput(145.34,92.49)(0.11,0.24){6}{\line(0,1){0.24}}
\multiput(146.01,93.94)(0.11,0.14){9}{\line(0,1){0.14}}
\multiput(146.98,95.21)(0.14,0.11){9}{\line(1,0){0.14}}
\multiput(148.22,96.22)(0.24,0.12){6}{\line(1,0){0.24}}
\multiput(149.65,96.92)(0.59,0.10){4}{\line(1,0){0.59}}
\multiput(9.00,91.00)(0.79,-0.09){2}{\line(1,0){0.79}}
\multiput(10.59,90.82)(0.30,-0.11){5}{\line(1,0){0.30}}
\multiput(12.09,90.28)(0.17,-0.11){8}{\line(1,0){0.17}}
\multiput(13.43,89.42)(0.11,-0.11){10}{\line(0,-1){0.11}}
\multiput(14.54,88.27)(0.12,-0.20){7}{\line(0,-1){0.20}}
\multiput(15.37,86.91)(0.10,-0.30){5}{\line(0,-1){0.30}}
\multiput(15.86,85.39)(0.07,-0.80){2}{\line(0,-1){0.80}}
\multiput(16.00,83.80)(-0.11,-0.79){2}{\line(0,-1){0.79}}
\multiput(15.77,82.22)(-0.12,-0.30){5}{\line(0,-1){0.30}}
\multiput(15.19,80.73)(-0.11,-0.16){8}{\line(0,-1){0.16}}
\multiput(14.29,79.42)(-0.13,-0.12){9}{\line(-1,0){0.13}}
\multiput(13.11,78.34)(-0.20,-0.11){7}{\line(-1,0){0.20}}
\multiput(11.72,77.55)(-0.38,-0.11){4}{\line(-1,0){0.38}}
\put(10.19,77.10){\line(-1,0){1.59}}
\multiput(8.60,77.01)(-0.52,0.09){3}{\line(-1,0){0.52}}
\multiput(7.03,77.28)(-0.25,0.10){6}{\line(-1,0){0.25}}
\multiput(5.56,77.90)(-0.16,0.12){8}{\line(-1,0){0.16}}
\multiput(4.27,78.84)(-0.12,0.13){9}{\line(0,1){0.13}}
\multiput(3.22,80.05)(-0.11,0.20){7}{\line(0,1){0.20}}
\multiput(2.48,81.46)(-0.10,0.39){4}{\line(0,1){0.39}}
\put(2.07,83.00){\line(0,1){1.60}}
\multiput(2.03,84.60)(0.11,0.52){3}{\line(0,1){0.52}}
\multiput(2.34,86.16)(0.11,0.24){6}{\line(0,1){0.24}}
\multiput(3.01,87.61)(0.11,0.14){9}{\line(0,1){0.14}}
\multiput(3.98,88.88)(0.14,0.11){9}{\line(1,0){0.14}}
\multiput(5.22,89.89)(0.24,0.12){6}{\line(1,0){0.24}}
\multiput(6.65,90.59)(0.59,0.10){4}{\line(1,0){0.59}}
\multiput(49.00,91.00)(0.79,-0.09){2}{\line(1,0){0.79}}
\multiput(50.59,90.82)(0.30,-0.11){5}{\line(1,0){0.30}}
\multiput(52.09,90.28)(0.17,-0.11){8}{\line(1,0){0.17}}
\multiput(53.43,89.42)(0.11,-0.11){10}{\line(0,-1){0.11}}
\multiput(54.54,88.27)(0.12,-0.20){7}{\line(0,-1){0.20}}
\multiput(55.37,86.91)(0.10,-0.30){5}{\line(0,-1){0.30}}
\multiput(55.86,85.39)(0.07,-0.80){2}{\line(0,-1){0.80}}
\multiput(56.00,83.80)(-0.11,-0.79){2}{\line(0,-1){0.79}}
\multiput(55.77,82.22)(-0.12,-0.30){5}{\line(0,-1){0.30}}
\multiput(55.19,80.73)(-0.11,-0.16){8}{\line(0,-1){0.16}}
\multiput(54.29,79.42)(-0.13,-0.12){9}{\line(-1,0){0.13}}
\multiput(53.11,78.34)(-0.20,-0.11){7}{\line(-1,0){0.20}}
\multiput(51.72,77.55)(-0.38,-0.11){4}{\line(-1,0){0.38}}
\put(50.19,77.10){\line(-1,0){1.59}}
\multiput(48.60,77.01)(-0.52,0.09){3}{\line(-1,0){0.52}}
\multiput(47.03,77.28)(-0.25,0.10){6}{\line(-1,0){0.25}}
\multiput(45.56,77.90)(-0.16,0.12){8}{\line(-1,0){0.16}}
\multiput(44.27,78.84)(-0.12,0.13){9}{\line(0,1){0.13}}
\multiput(43.22,80.05)(-0.11,0.20){7}{\line(0,1){0.20}}
\multiput(42.48,81.46)(-0.10,0.39){4}{\line(0,1){0.39}}
\put(42.07,83.00){\line(0,1){1.60}}
\multiput(42.03,84.60)(0.11,0.52){3}{\line(0,1){0.52}}
\multiput(42.34,86.16)(0.11,0.24){6}{\line(0,1){0.24}}
\multiput(43.01,87.61)(0.11,0.14){9}{\line(0,1){0.14}}
\multiput(43.98,88.88)(0.14,0.11){9}{\line(1,0){0.14}}
\multiput(45.22,89.89)(0.24,0.12){6}{\line(1,0){0.24}}
\multiput(46.65,90.59)(0.59,0.10){4}{\line(1,0){0.59}}
\multiput(29.00,106.00)(0.79,-0.09){2}{\line(1,0){0.79}}
\multiput(30.59,105.82)(0.30,-0.11){5}{\line(1,0){0.30}}
\multiput(32.09,105.28)(0.17,-0.11){8}{\line(1,0){0.17}}
\multiput(33.43,104.42)(0.11,-0.11){10}{\line(0,-1){0.11}}
\multiput(34.54,103.27)(0.12,-0.20){7}{\line(0,-1){0.20}}
\multiput(35.37,101.91)(0.10,-0.30){5}{\line(0,-1){0.30}}
\multiput(35.86,100.39)(0.07,-0.80){2}{\line(0,-1){0.80}}
\multiput(36.00,98.80)(-0.11,-0.79){2}{\line(0,-1){0.79}}
\multiput(35.77,97.22)(-0.12,-0.30){5}{\line(0,-1){0.30}}
\multiput(35.19,95.73)(-0.11,-0.16){8}{\line(0,-1){0.16}}
\multiput(34.29,94.42)(-0.13,-0.12){9}{\line(-1,0){0.13}}
\multiput(33.11,93.34)(-0.20,-0.11){7}{\line(-1,0){0.20}}
\multiput(31.72,92.55)(-0.38,-0.11){4}{\line(-1,0){0.38}}
\put(30.19,92.10){\line(-1,0){1.59}}
\multiput(28.60,92.01)(-0.52,0.09){3}{\line(-1,0){0.52}}
\multiput(27.03,92.28)(-0.25,0.10){6}{\line(-1,0){0.25}}
\multiput(25.56,92.90)(-0.16,0.12){8}{\line(-1,0){0.16}}
\multiput(24.27,93.84)(-0.12,0.13){9}{\line(0,1){0.13}}
\multiput(23.22,95.05)(-0.11,0.20){7}{\line(0,1){0.20}}
\multiput(22.48,96.46)(-0.10,0.39){4}{\line(0,1){0.39}}
\put(22.07,98.00){\line(0,1){1.60}}
\multiput(22.03,99.60)(0.11,0.52){3}{\line(0,1){0.52}}
\multiput(22.34,101.16)(0.11,0.24){6}{\line(0,1){0.24}}
\multiput(23.01,102.61)(0.11,0.14){9}{\line(0,1){0.14}}
\multiput(23.98,103.88)(0.14,0.11){9}{\line(1,0){0.14}}
\multiput(25.22,104.89)(0.24,0.12){6}{\line(1,0){0.24}}
\multiput(26.65,105.59)(0.59,0.10){4}{\line(1,0){0.59}}
\multiput(29.00,81.00)(0.79,-0.09){2}{\line(1,0){0.79}}
\multiput(30.59,80.82)(0.30,-0.11){5}{\line(1,0){0.30}}
\multiput(32.09,80.28)(0.17,-0.11){8}{\line(1,0){0.17}}
\multiput(33.43,79.42)(0.11,-0.11){10}{\line(0,-1){0.11}}
\multiput(34.54,78.27)(0.12,-0.20){7}{\line(0,-1){0.20}}
\multiput(35.37,76.91)(0.10,-0.30){5}{\line(0,-1){0.30}}
\multiput(35.86,75.39)(0.07,-0.80){2}{\line(0,-1){0.80}}
\multiput(36.00,73.80)(-0.11,-0.79){2}{\line(0,-1){0.79}}
\multiput(35.77,72.22)(-0.12,-0.30){5}{\line(0,-1){0.30}}
\multiput(35.19,70.73)(-0.11,-0.16){8}{\line(0,-1){0.16}}
\multiput(34.29,69.42)(-0.13,-0.12){9}{\line(-1,0){0.13}}
\multiput(33.11,68.34)(-0.20,-0.11){7}{\line(-1,0){0.20}}
\multiput(31.72,67.55)(-0.38,-0.11){4}{\line(-1,0){0.38}}
\put(30.19,67.10){\line(-1,0){1.59}}
\multiput(28.60,67.01)(-0.52,0.09){3}{\line(-1,0){0.52}}
\multiput(27.03,67.28)(-0.25,0.10){6}{\line(-1,0){0.25}}
\multiput(25.56,67.90)(-0.16,0.12){8}{\line(-1,0){0.16}}
\multiput(24.27,68.84)(-0.12,0.13){9}{\line(0,1){0.13}}
\multiput(23.22,70.05)(-0.11,0.20){7}{\line(0,1){0.20}}
\multiput(22.48,71.46)(-0.10,0.39){4}{\line(0,1){0.39}}
\put(22.07,73.00){\line(0,1){1.60}}
\multiput(22.03,74.60)(0.11,0.52){3}{\line(0,1){0.52}}
\multiput(22.34,76.16)(0.11,0.24){6}{\line(0,1){0.24}}
\multiput(23.01,77.61)(0.11,0.14){9}{\line(0,1){0.14}}
\multiput(23.98,78.88)(0.14,0.11){9}{\line(1,0){0.14}}
\multiput(25.22,79.89)(0.24,0.12){6}{\line(1,0){0.24}}
\multiput(26.65,80.59)(0.59,0.10){4}{\line(1,0){0.59}}
\put(9.00,84.00){\makebox(0,0)[cc]{$U_0$}}
\put(49.00,84.00){\makebox(0,0)[cc]{$U_{12}$}}
\put(29.00,99.00){\makebox(0,0)[cc]{$U_2$}}
\put(29.00,74.00){\makebox(0,0)[cc]{$U_1$}}
\put(42.00,73.39){\makebox(0,0)[cc]{$\theta_1$}}
\put(17.00,73.39){\makebox(0,0)[cc]{$\omega_1$}}
\put(16.33,95.79){\makebox(0,0)[cc]{$\omega_2$}}
\put(40.00,95.79){\makebox(0,0)[lc]{$\theta_2 = -
\frac{\omega_1\theta_1}{\omega_2}$}}
\put(82.00,70.33){\makebox(0,0)[cc]{$U_0$}}
\put(132.00,70.33){\makebox(0,0)[cc]{$U_1$}}
\put(132.00,120.33){\makebox(0,0)[cc]{$U_{13}$}}
\put(102.00,140.33){\makebox(0,0)[cc]{$U_{23}$}}
\put(152.00,140.33){\makebox(0,0)[cc]{$U_{123}$}}
\put(102.00,90.33){\makebox(0,0)[cc]{$U_2$}}
\put(152.00,90.33){\makebox(0,0)[cc]{$U_{12}$}}
\put(146.67,77.12){\makebox(0,0)[cc]{$\omega _1'$}}
\put(123.00,97.12){\makebox(0,0)[cc]{$\omega _2'$}}
\put(89.33,87.52){\makebox(0,0)[cc]{$\omega _2$}}
\put(107.00,63.52){\makebox(0,0)[cc]{$\omega _1$}}
\put(76.33,95.52){\makebox(0,0)[cc]{$\omega_3$}}
\put(97.00,110.33){\makebox(0,0)[cc]{$\theta_2$}}
\put(136.67,101.12){\makebox(0,0)[cc]{$\theta_1$}} 
\put(158.00,115.52){\makebox(0,0)[cc]{$\theta'$}}
\put(117.00,45.33){\makebox(0,0)[cc]{~~~}}
\put(29.00,44.00){\makebox(0,0)[cc]{~~~}}
\put(23.67,94.19){\line(-5,-3){9.67}}
\put(15.33,81.39){\line(2,-1){7.67}}
\put(34.67,77.39){\line(5,2){8.00}}
\put(42.67,87.79){\line(-4,3){8.33}}
\put(82.00,113.92){\line(0,-1){36.80}}
\put(89.00,70.72){\line(1,0){36.00}}
\put(145.00,90.72){\line(-1,0){35.67}}
\put(102.00,97.12){\line(0,1){36.80}}
\put(109.00,140.32){\line(1,0){36.00}}
\put(152.00,133.92){\line(0,-1){36.80}}
\put(132.00,77.12){\line(0,1){36.80}}
\put(125.00,120.32){\line(-1,0){36.00}}
\put(147.00,135.33){\line(-1,-1){10.67}}
\put(137.33,75.52){\line(1,1){10.33}}
\put(97.33,85.12){\line(-6,-5){10.67}}
\put(97.33,134.72){\line(-5,-4){11.00}}
\put(82.33,120.33){\makebox(0,0)[cc]{$U_3$}}
\end{picture}
\end{center}
\caption{Rhombic and cubic superposition formulas for
$(2+1)$-dimensional integrable systems} \label{fig-cube}
\end{figure}
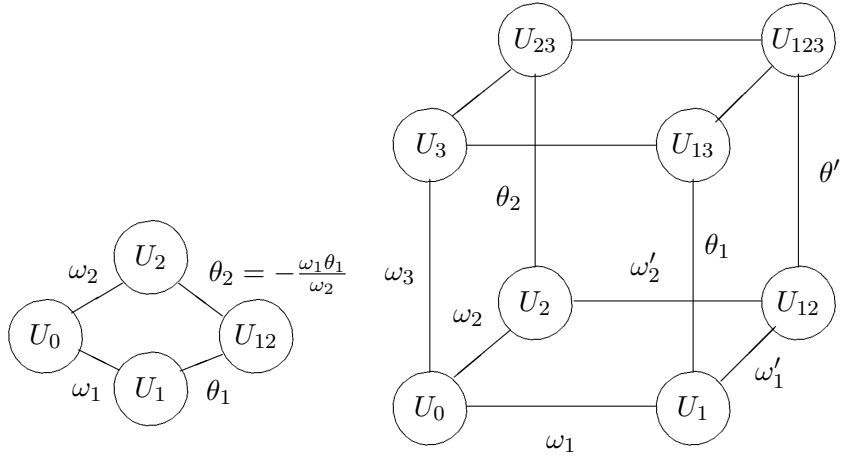

Namely, as one may check, the following statement, obtained by
L.~Bi\-an\-chi~\cite{bianchi} for the case of the Moutard equation
$\varphi_{xy}= u(x,y)\varphi$, is also valid in our case:
\begin{theorem}\label{th-cube}
If $\omega_1$, $\omega_2$, $\omega_3$ are three solution of
 (\ref{eq}) where $U=U_0$, $V=V_0$, $V^*=V^*_0$ is
some given solution of the Novikov-Veselov equation (\ref{nv}), and
$\theta_1$, $\theta_2$ (transformed solutions of (\ref{eq}) with new
$U_1= U_0 - 2\partial\bar\partial(\ln \omega_1)$, $U_2= U_0 -
2\partial\bar\partial(\ln \omega_2)$ and respectively transformed
$V_i$, $V^*_i$---another set of solutions of the Novikov-Veselov
equation) are obtained from $\omega_3$ via (\ref{ext}) then there
exists a unique solution $\theta'$ of the 4th linear system
(\ref{eq}) with the potentials $U=U_{12}$, $V=V_{12}$,
$V^*=V^*_{12}$ such that $\theta'$ is connected to $\theta_1$,
$\theta_2$ with the extended Moutard transformations (\ref{ext}).
This $\theta'$ is expressible with an algebraic formula
\begin{equation}\label{bi-cu}
\theta' - \omega_3= \frac{\omega_1 \omega_2}{\lambda}(\theta_2 -
\theta_1), \qquad \lambda = \omega_1 \omega_1'= - \omega_2
\omega_2',
\end{equation}
where $\omega_1'$, $\omega_2'$ are obtained from $\omega_1$,
$\omega_2$ according to (\ref{ext}).
\end{theorem}
It is useful to explain this statement and the ``extended
superposition formula'' (\ref{ext}) by commutative rhombic and cubic
diagrams (``Bianchi superposition cube'') shown on
Fig.~\ref{fig-cube}.

The rhombic diagram in the left part of Fig.~\ref{fig-cube} shows
two solutions $\omega_1$ and $\omega_2$ of (\ref{eq}) with the
initial potential $U_0$ and the respective $V_0$, $V^*_0$ (which
form a solution of the Novikov-Veselov equation (\ref{nv})). Those
$\omega_i$ produce two new potentials $U_1$,  $U_2$ (and the
respective $V_i$, $V^*_i$)--- two new solutions of the
Novikov-Veselov equation (this is \emph{step one}). The formula
(\ref{ext}) then gives us (a set of) $\theta_1$---a solution of
(\ref{eq}) with the potentials $U_1$, $V_1$, $V^*_1$ as well as
$\theta_2  = -\frac{\omega_1}{\omega_2} \theta_1$---a solution of
(\ref{eq}) with the potentials $U_2$, $V_2$, $V^*_2$. Using either
$\theta_1$ or $\theta_2$ we can find the potentials $U_{12}=U_1 +
2\partial\bar{\partial}\log \theta_1=U_2 +
2\partial\bar{\partial}\log \theta_2$, $V_{12}=V_1 + 2\partial^2\log
\theta_1=V_2 + 2\partial^2\log \theta_2$, $V^*_{12}=V^*_1 +
2\bar{\partial}^2\log \theta_1=V^*_2 + 2\bar{\partial}^2\log
\theta_2$ shown as the last fourth circle of the rhombic diagram
(this is \emph{step two}). As we see, this second step requires a
quadrature and produces a free integration constant in the final
potentials $U_{12}$, $V_{12}$, $V^*_{12}$.

If we have done those two steps for each of the three pairs
$\{\omega_1,\omega_2\}$, $\{\omega_1,\omega_3\}$,
$\{\omega_2,\omega_3\}$ shown on the right cubic diagram (and we
have accumulated also three additional constants of integration),
the Bianchi algebraic superposition formula (\ref{bi-cu}) gives a
solution $\theta'$ of (\ref{eq}) for the set $U_{12}$, $V_{12}$,
$V^*_{12}$ of the potentials obtained on step two.

The formula (\ref{bi-cu}) was proved in \cite{bianchi} for the case
of the Moutard equation $\omega_{xy}=U(x,y)\omega$. A
straightforward computation shows that it preserves the dynamics of
the potentials defined by (\ref{nv}), as Theorem~\ref{th-cube}
states.

Using (\ref{eq}) and (\ref{bi-cu}) one can see that

\begin{itemize}
\item
{\sl if we start with $U_0=V_0=V^*=0$ and take arbitrary harmonic
polynomials as the initial solutions $\omega_i$, $i=1,\ldots, N$,
then the potentials obtained on the first two steps are rational
functions of $(x,y,t)$. 
All subsequent potentials obtained on the next steps will be
rational solutions of the Novikov-Veselov equation as well.}
\item
{\sl If we know some large set $\{\psi\}$ of solutions of the
Schr\"odinger equation $(\partial\bar\partial +U_0)\psi=0$ for the
initial potential $U_0$, then using Theorem~\ref{th-cube} we can
construct the corresponding solutions of $(\partial\bar\partial
+U_{12})\tilde\psi=0$ using quadratures (on step two) if we simply
set $\omega_3=\psi$ in the superposition formulas (\ref{eq}) and
(\ref{bi-cu}).}
\item
{\sl This proves that we obtain potentials $U_{12}$ integrable on
zero energy level if we start from some integrable $U_0$, for
example $U_0=0$.}
\end{itemize}

\section{A remark on
periodic and quasiperiodic solitons}\label{sec-per}

The similar construction can be applied to the case when the initial functions
$\omega_1$ and $\omega_2$ satisfy the Schr\"odinger equation
with the constant potential $u = -k^2$:
$$
(-\Delta - k^2)\omega_1 =  (-\Delta - k^2)\omega_2 = 0.
$$
It appears that the second iteration may give an integrable potential which
is periodic and smooth. Let us expose one such an example.

Let
$$
\omega_1 = \sin kx, \ \ \
\omega_2 = \sin(ax+by), \ a^2+b^2=k^2,
$$
and the Moutard transformation defined by $\omega_1$ maps $\omega_2$
into $\theta_1$ (modulo multiples of $\frac{1}{\omega_1}$):
$$
\theta_1 = \frac{b}{2 \sin kx} \left( \frac{\cos(ax+by+kx)}{a+k} -
\frac{\cos(ax+by-kx)}{a-k}+ C\right)
$$
with $C = \mathrm{const}$ and
the new potential is
$$
\widetilde{u} = k^2 +2\frac{\cos^2 kx}{\sin^2 kx}.
$$
Let us iterate the Moutard transformation using $\theta_1$ as the
generating solution and obtain a new potential
$$
\widetilde{\widetilde{u}} = -\widetilde{u}+ 2\frac{(\theta_1)_x^2 +
(\theta_1)_y^2}{\theta_1^2}=k^2 -2\Delta\log(\omega_1\theta_1)
$$
which is nonsingular for $C$ large enough. An example of such a
periodic  potential $\widetilde{\widetilde{u}}$ and the
corresponding solution $\psi_1=1/\theta_1$ for the constants $C=3$,
$a=0$, $b=1$, $k=1$ is shown on Fig.~8,~9, page~\pageref{page-pic2}.

As we see by iterating the Moutard transformation one may construct
integrable two-dimensional non-singular periodic and quasiperiodic
(when   $a/b$ is not rational) potentials. The complete basis of
solutions for the constructed potentials
$\widetilde{\widetilde{u}}=-4U_{12}$ may be obtained by applying
(\ref{bi-cu}) with $\omega_3=\exp\big(i(px+qy)\big)$, $p^2+q^2=k^2$.
The spectral properties and the NV evolution of such potentials
deserve an additional study.

\newpage

\label{page-pic}

\unitlength 0.5mm \linethickness{0.5pt}
\begin{picture}(64.67,78.00)
\put(-50.67,17.00){\includegraphics[width=0.60\textwidth]{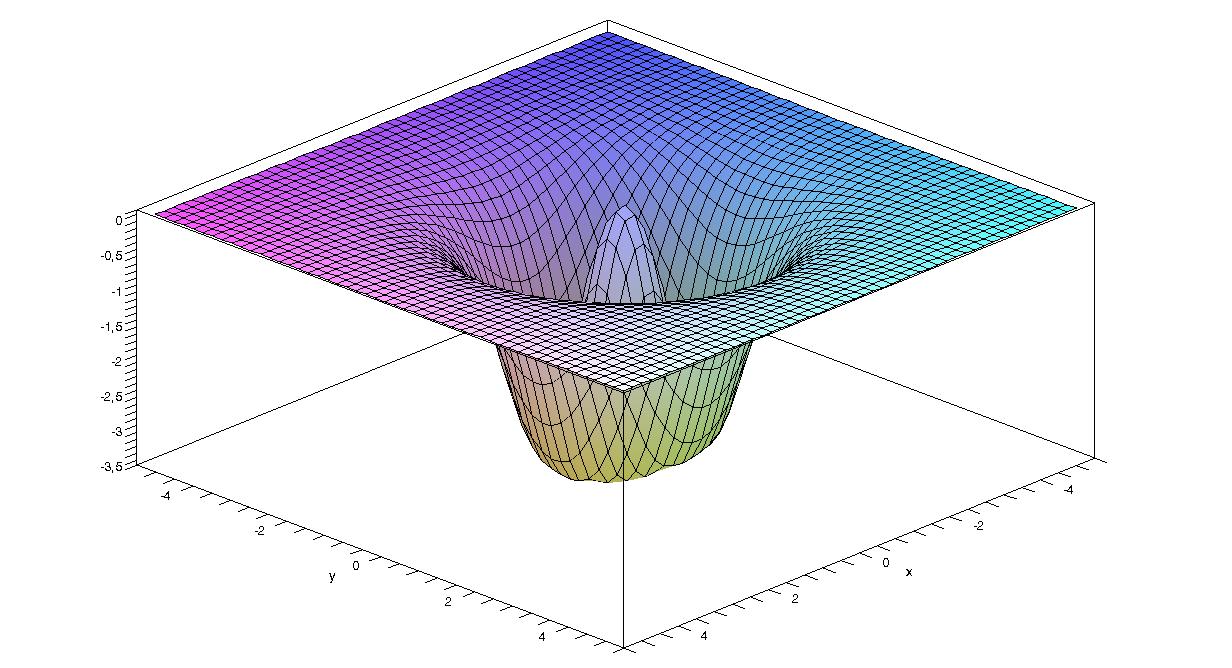}}
\put(10.67,5.00){\makebox(0,0)[cc]{Figure 2: The potential
(\ref{ord2-1}).}}
\put(110.67,17.00){\includegraphics[width=0.60\textwidth]{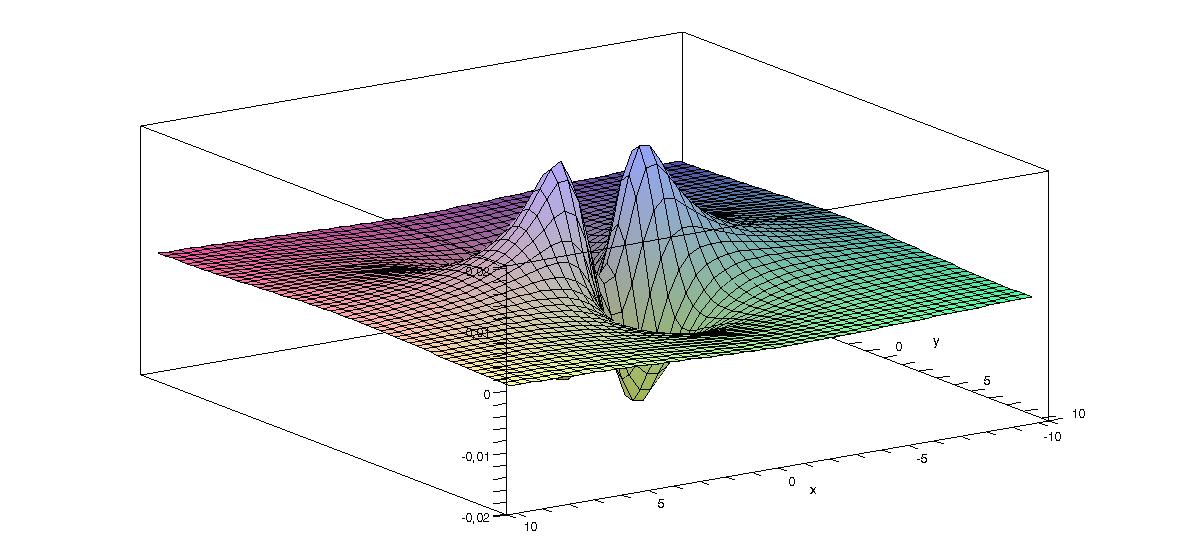}}
\put(160.67,5.00){\makebox(0,0)[cc]{Figure 3: The solution $\psi_1$
in (\ref{ord2-2}).}}
\end{picture}

\unitlength 0.5mm \linethickness{0.5pt}
\begin{picture}(64.67,128.00)
\put(-50.67,17.00){\includegraphics[width=0.60\textwidth]{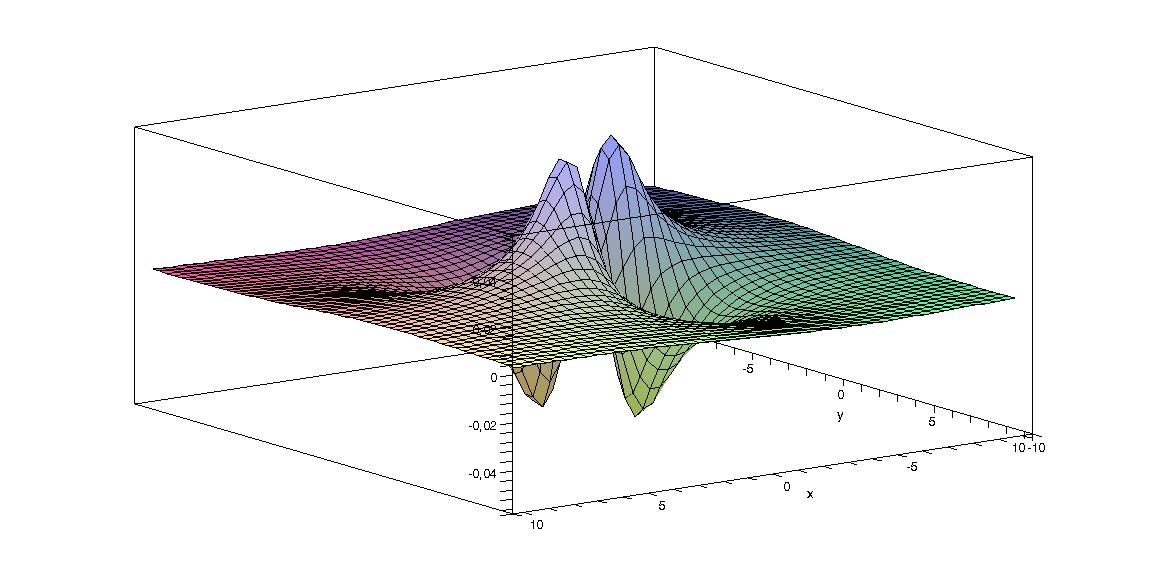}}
\put(10.67,5.00){\makebox(0,0)[cc]{Figure 4: The solution $\psi_2$
in (\ref{ord2-2}).}}
\end{picture}

\unitlength 0.5mm \linethickness{0.5pt}
\begin{picture}(64.67,118.00)
\put(-50.67,17.00){\includegraphics[width=0.60\textwidth]{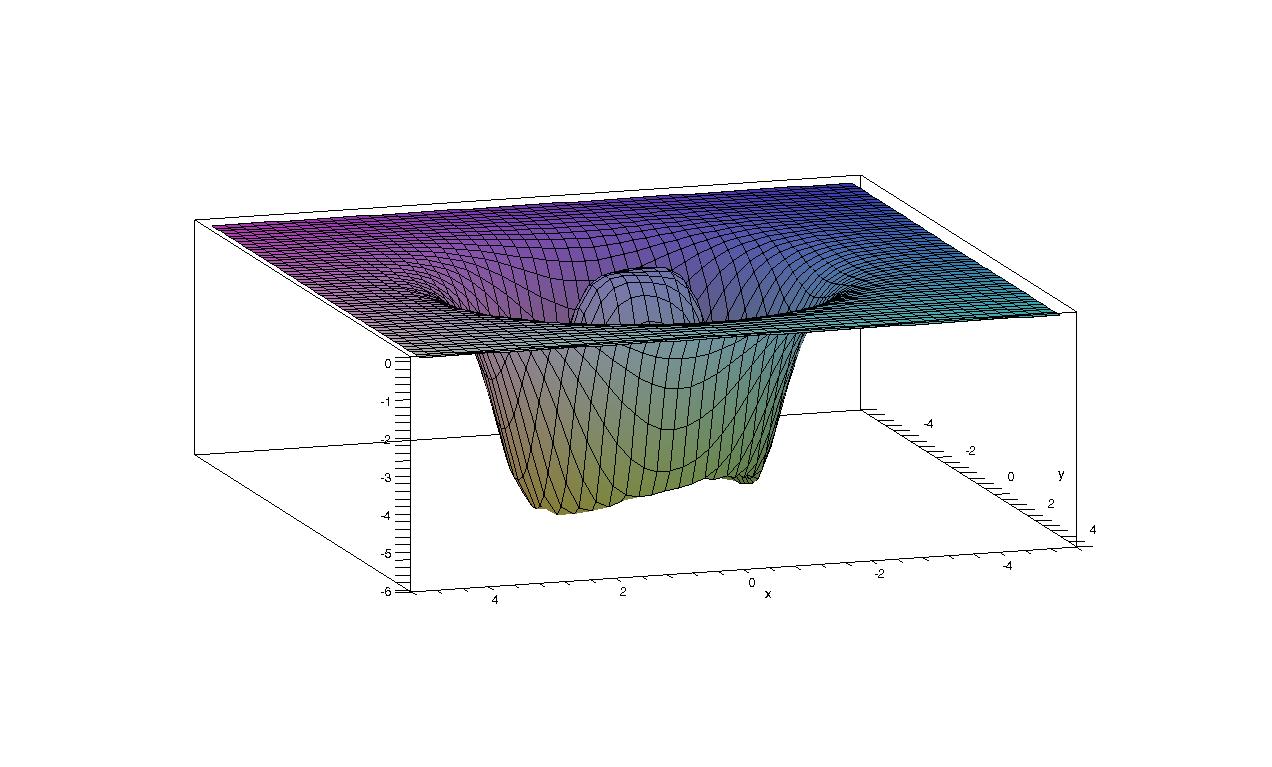}}
\put(10.67,5.00){\makebox(0,0)[cc]{Figure 5: The potential $u$ in
(\ref{ord3}).}}
\put(110.67,17.00){\includegraphics[width=0.60\textwidth]{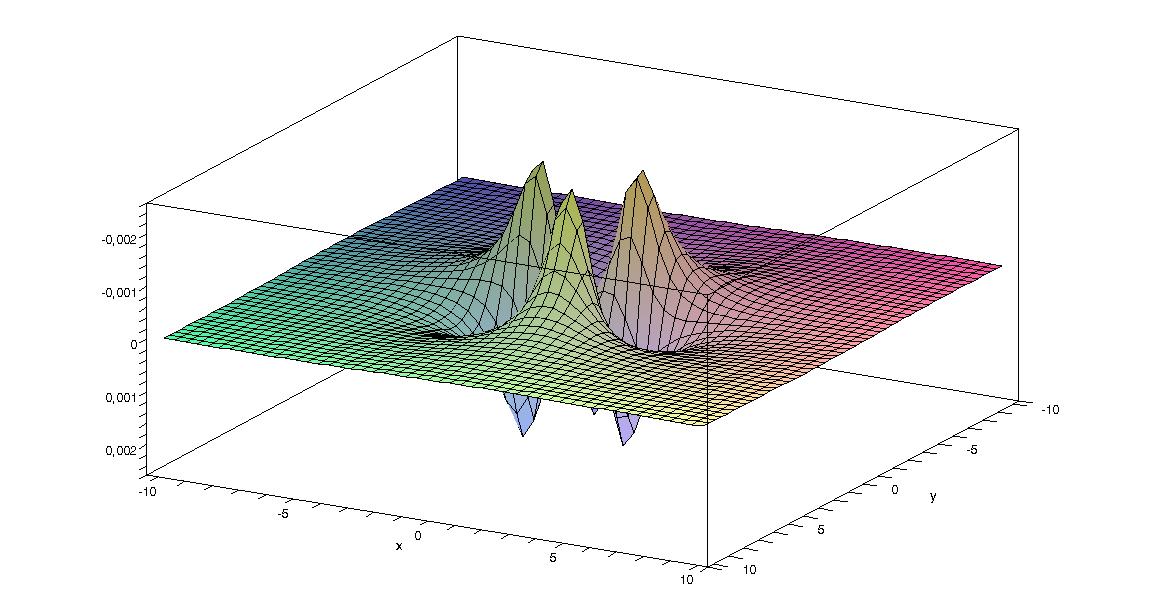}}
\put(160.67,5.00){\makebox(0,0)[cc]{Figure 6: The solution $\psi_1$
in (\ref{ord3}).}}
\end{picture}

\newpage

\unitlength 0.5mm \linethickness{0.5pt}
\begin{picture}(64.67,68.00)
\put(-50.67,17.00){\includegraphics[width=0.60\textwidth]{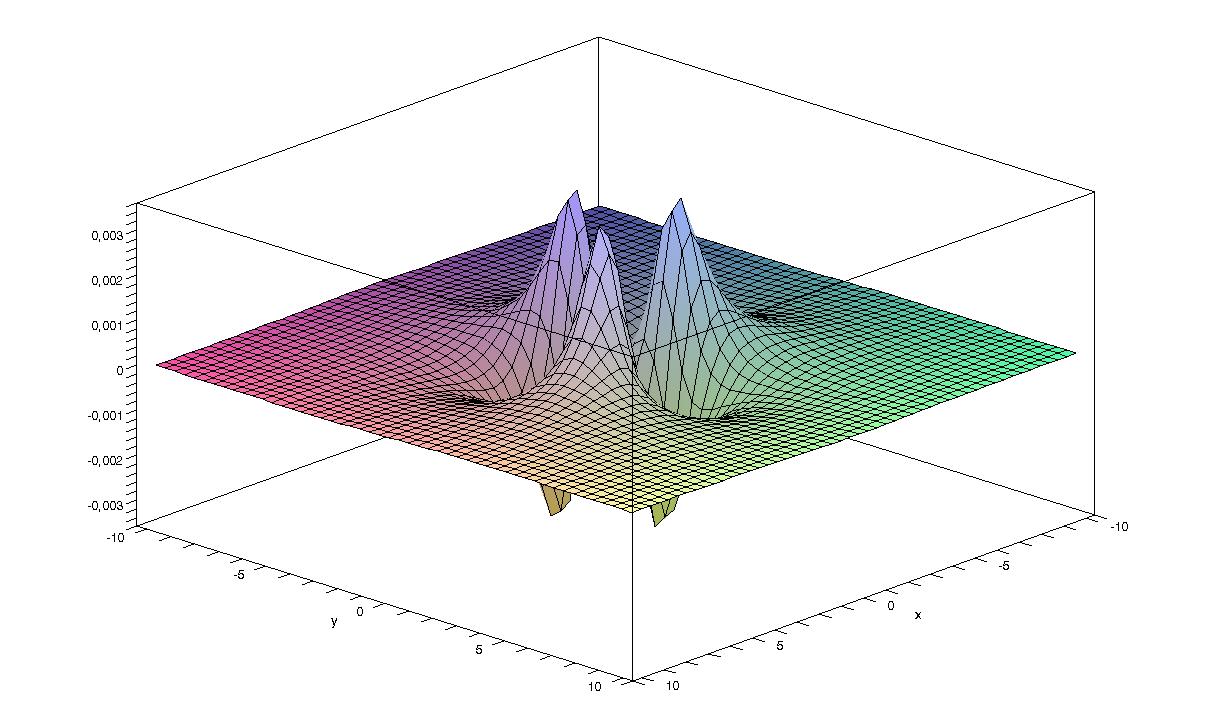}}
\put(10.67,5.00){\makebox(0,0)[cc]{Figure 7: The solution $\psi_2$
in (\ref{ord3}).}}
\end{picture}

\label{page-pic2}

\unitlength 0.5mm \linethickness{0.5pt}
\begin{picture}(64.67,128.00)
\put(-50.67,17.00){\includegraphics[width=0.60\textwidth]{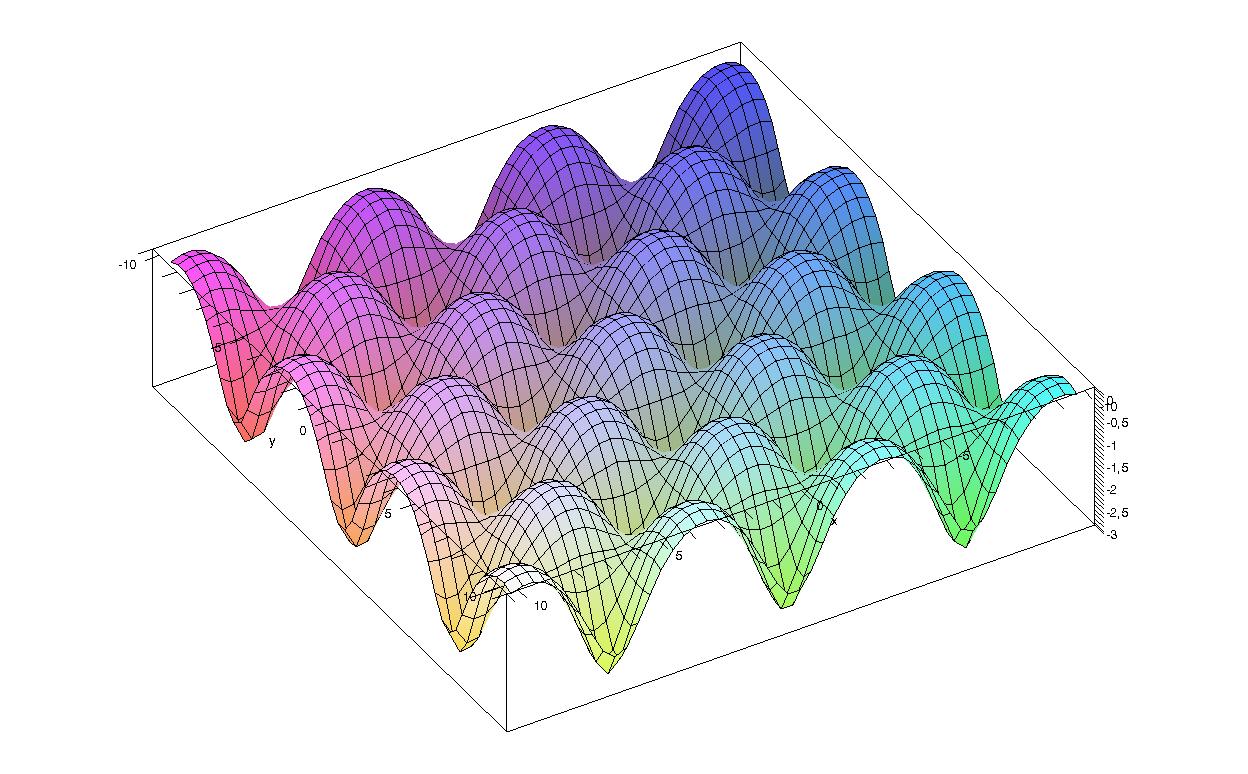}}
\put(10.67,5.00){\makebox(0,0)[cc]{Figure 8: The periodic  potential
$\widetilde{\widetilde{u}}$, Section~\ref{sec-per}.}}
\end{picture}

\unitlength 0.5mm \linethickness{0.5pt}
\begin{picture}(64.67,128.00)
\put(-50.67,17.00){\includegraphics[width=0.60\textwidth]{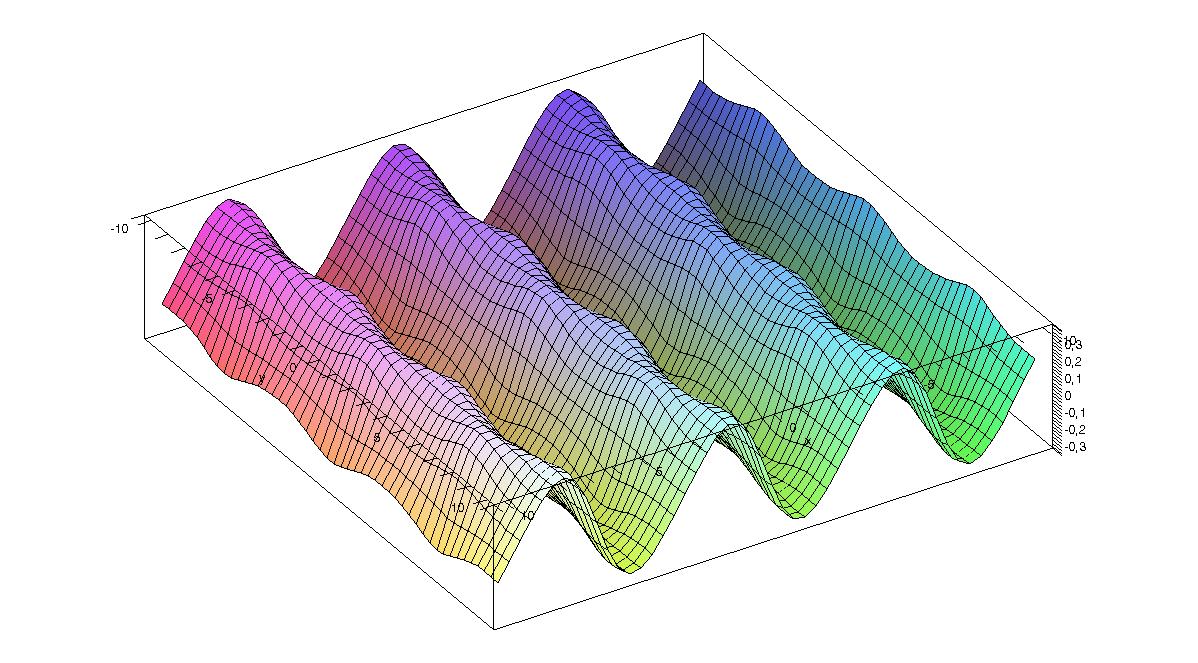}}
\put(40.67,5.00){\makebox(0,0)[cc]{Figure 9: The solution $\psi_1$
for this periodic potential, Section~\ref{sec-per}.}}
\end{picture}

\end{document}